\documentclass[11pt]{iopart}
\usepackage{bm}
\usepackage{hyperref}
\usepackage{amssymb}
\usepackage{feynmp}
\usepackage{amsopn}
\usepackage{setstack}
\RequirePackage{ifpdf}

\ifpdf
\else % ordinary latex seems to include these as eps files without a problem
\fi

\eqnobysec

\renewcommand{\etal}{\emph{et al.}}

\newcommand{\R}{\mathcal{R}}
\newcommand{\fnl}{f_{\mathrm{NL}}}
\newcommand{\Ps}{\mathcal{P}}
\newcommand{\Planck}{M_{\mathrm{P}}}
\newcommand{\bra}[1]{\langle{{#1}}|}
\newcommand{\ket}[1]{|{{#1}}\rangle}

\newcommand{\gdiff}{\bar{\partial}}
\newcommand{\EulerMascheroni}{\gamma_{\mathrm{E}}}

\renewcommand{\d}{\mathrm{d}}
\newcommand{\vect}[1]{\bm{\mathrm{{#1}}}}
\renewcommand{\e}[1]{\mathrm{e}^{{#1}}}
\newcommand{\im}{\mathrm{i}}

\newcommand{\grad}{\nabla}

\renewcommand{\geq}{\geqslant}

\newcommand{\twoEi}[4]{{\left[\hspace{-1.33mm}\left[%
  \scriptsize%
  \begin{array}{@{\hspace{0mm}}c@{\hspace{2mm}}c@{\hspace{0mm}}}%
    {#1} & {#2} \\%
    {#3} & {#4}%
  \end{array}\right]\hspace{-1.33mm}\right]}}
\newcommand{\threeEi}[6]{{\left[\hspace{-1.3mm}\left[%
  \scriptsize%
  \begin{array}{@{\hspace{0mm}}c@{\hspace{2mm}}c@{\hspace{2mm}}c%
	@{\hspace{0mm}}}%
    {#1} & {#2} & {#3} \\%
    {#4} & {#5} & {#6}%
  \end{array}\right]\hspace{-1.33mm}\right]}}
\newcommand{\threeEiVariant}[6]{{\left|\hspace{-1.25mm}\left[%
  \scriptsize%
  \begin{array}{@{\hspace{0mm}}c@{\hspace{2mm}}c@{\hspace{2mm}}c%
	@{\hspace{0mm}}}%
    {#1} & {#2} & {#3} \\%
    {#4} & {#5} & {#6}%
  \end{array}\right]\hspace{-1.25mm}\right|}}

\makeatletter
\newcommand\numberwithin[2]{\@addtoreset{#1}{#2}}
\makeatother
\numberwithin{footnote}{section}

\begin{document}
\begin{fmffile}{diags}
	\title{Magnetogenesis and the primordial non-gaussianity}
	
	\author{David Seery}
	
	\address{Department of Applied Mathematics and Theoretical Physics, \\
	Wilberforce Road, Cambridge, CB3 0WA, United Kingdom}

	\eads{\mailto{djs61@cam.ac.uk}}

	\begin{abstract}
		The primordial density fluctuation
		inevitably couples to all forms of matter via loop corrections
		and depends on the
		ambient conditions while inflation was ongoing. This gives us
		an opportunity to observe
		processes which were in progress
		while the universe was inflating, provided they were sufficiently
		dramatic to overcome suppression by powers of
		$(H/\Planck)^2 \approx 10^{-9}$,
		where $H$ is the Hubble scale during inflation and
		$\Planck$ is the Planck mass.
		As an example, if a primordial magnetic field was synthesized during
		inflation, as suggested by some interpretations of the apparently
		universal $10^{-6}$ gauss field observed on galactic scales,
		then this could leave traces in inflationary observables.
		In this paper, I compute corrections to the spectrum and
		bispectrum generated by a varying electromagnetic coupling
		during inflation, assuming that the variation in this coupling is
		mediated by interaction with a collection of light scalar fields.
		If the mass scale associated with this interaction is too far
		below the Planck scale then the stability of perturbation theory
		can be upset. For the mass-scale
		which is relevant in the standard magnetogenesis scenario, however,
		the theory is stable and the model is apparently consistent with
		observational constraints.
				
	\vspace{3mm}
	\begin{flushleft}
	\end{flushleft}
	\end{abstract}
	\maketitle

	\section{Introduction}
	\label{sec:introduction}
	
	Our understanding of the very early universe has settled down over the
	last several years to give, on balance, a broadly consistent picture.
	In this
	\emph{concordance} model, an adiabatic perturbation is synthesized
	on superhorizon scales during an early epoch of inflation and is
	converted by gravitational collapse into the observed distribution
	of baryons and cold dark matter in the universe.
	It is generally accepted that this simple timeline is sufficient to
	explain the vast majority of observational data.
	
	A zoo of
	optional components can be added to improve the fit to certain small
	anomalies in the data, or to provide an origin for features of our
	macroscopic world which are presently without a theoretical rationale.
	Examples of these optional features are scalar isocurvature modes
	(which may source non-gaussianities in the temperature anistropy
	\cite{Vernizzi:2006ve,Battefeld:2006sz}),
	vector modes (which may source statistical anisotropy
	\cite{Yokoyama:2008xw,Kanno:2008gn,Dimopoulos:2008yv}),
	or tensor gravitational waves.
	All of these must be subdominant to the adiabatic perturbation.
	We usually assume
	that these optional extras can be
	added or taken away without penalty.
	However, if they mediate sufficiently
	dramatic processes then
	we must remember that the observable
	adiabatic fluctuation will couple to all of them via loop corrections.
	In this paper the loop corrections arising from an especially interesting
	optional addition are studied, namely a varying electromagnetic%
		\footnote{In fact, the coupling in question cannot be the
		electromagnetic coupling $\alpha$, the fine structure ``constant'',
		because inflation is usually supposed to take place at an
		energy scale far above the scale of electroweak symmetry breaking,
		and at such energies the electromagnetic field has not yet obtained
		a separate identity.
		
		It is known that non-Abelian gauge fields are
		shielded from obtaining a perturbation during inflation by their
		strong self-interactions \cite{Davis:1995mv},
		and therefore the gauge coupling in
		question must be the $U(1)$ Standard Model hypercharge field. A
		fluctuation in this field will communicate some perturbation to
		the electromagnetic field at electroweak symmetry breaking.
		However, the detailed identity of the Abelian gauge field
		in question is irrelevant for the question posed in the present
		paper. For simplicity, I will refer to this field as the
		``electromagnetic'' field and its coupling as the
		``electromagnetic coupling'' throughout, with the
		understanding that for applications to magnetogenesis
		the hypercharge field must be substituted and the resulting
		fluctuation projected onto the physical electromagnetic field.
		Similarly, this analysis will apply to any $U(1)$ fields in the
		low energy theory of inflation whatever their
		microscopic origin, although not to any non-Abelian fields.}
	coupling which is mediated by interaction with
	a collection of light scalar
	fields. Among other effects, a varying coupling of this sort could be
	responsible
	for generating a primordial magnetic field. Such magnetic fields are
	required by observation \cite{Han:2002ns}
	and although a number of possible formation
	mechanisms are known---see, for example, Ref.~\cite{Battefeld:2007qn}
	for a non-inflationary example---%
	their large-scale coherence and homogeneity
	hints at an inflationary origin parallel with the origin of the
	primordial density fluctuation.
	
	What can we hope to learn from the study of such loop effects?
	One motivation is that, despite increasing high-quality
	experimental data which suggests the simplest predictions of
	inflation are a good fit for observation,
	it is still unclear whether inflation is the right model of microphysics
	or simply a good parametrization of an approximately
	gaussian, scale-invariant and
	adiabatic spectrum. One way to approach this question is to study
	the leading departures from gaussianity, as measured by the bispectrum
	and trispectrum.
	Non-gaussian statistics have received considerable attention
	over the last several years.
	Indeed, recent experimental results hint that it may
	be possible to discriminate among different microphysical
	models using this technique \cite{Yadav:2007yy,Komatsu:2008hk}.
	Such studies fall into
	the general framework of non-linear perturbation theory.
	However,
	departures of the observable adiabatic fluctuation
	from exact gaussianity are not the only effect we can hope
	to probe using this tool. In a quantum mechanical world which includes
	gravity we must expect the adiabatic fluctuation to couple to
	all other degrees of freedom which were light enough to be
	excited during the inflationary era.
	This point has been emphasized in recent work by Weinberg
	\cite{Weinberg:2008nf}. If
	inflation does more for us than merely generate the observed density
	fluctuation---and if we
	are lucky---it may be possible to probe whatever other processes were
	in progress during inflation, provided they were sufficiently
	dramatic to leave traces in the inflationary observables.
	The analysis given in this paper can be thought of as an exploratory
	step in this direction.
	
	There are other motivations. If we are to have confidence in our
	predictions, it is necessary to maintain control over the
	theoretical tools which we use.
	This will become of increasing significance as the data
	improve and we pass from a qualitative to a quantitative description
	of the earliest times, excluding some models as possible theories of
	the early universe while promoting others as a better match for
	observation. The key
	criterion here is stability of the perturbative
	series which is used to extract observables from the
	Lagrangian, a
	question which has already attracted
	attention in the literature \cite{Leblond:2008gg,ArmendarizPicon:2008yv}.
	Several potential sources of instability exist.
	Increasing orders of perturbation theory are typically suppressed
	by powers of the ratio $(H/\Planck)^2 \approx 10^{-9}$,
	where $H$ is the energy scale of inflation and $\Planck$ is
	the Planck mass. However, perturbation theory can in principle contain
	instabilities which scale like a positive power of
	the scale factor $a(t) \approx \exp(Ht)$ during inflation.
		
	If such ``fast''
	divergences are present then they will rapidly overwhelm any
	powers of $(H/\Planck)^2$ and render perturbation theory unstable after
	a short time---generally too short to be of any use in extracting
	predictions for relics of the early universe which are visible at the
	present day. Alternatively there may be large corrections which
	come from a sensitivity to physics in the ultra-violet, or from
	some other hierarchy which exists in the theory. The first possibility
	was studied by Armendariz-Picon {\etal} \cite{ArmendarizPicon:2008yv},
	whereas an example of the latter,
	which was studied by Leblond \& Shandera \cite{Leblond:2008gg},
	is the relative hierarchy $c_{\mathrm{s}}^{-2} \geq 1$ between the speed
	of sound and the speed of light. Whatever the source of large hierarchies
	which overwhelms the smallness of $(H/\Planck)^2$,
	it is important to emphasize that an instability in perturbation
	theory does not necessarily imply that anything untoward is taking
	place. It may simply mean that we need to find a better description
	of the process in question.
	
	The possibility of fast instabilities in perturbation
	theory was considered by Weinberg \cite{Weinberg:2005vy,
	Weinberg:2006ac} (see also Chaicherdsakul \cite{Chaicherdsakul:2006ui}),
	who gave a criterion according to
	which it is possible to decide whether such instabilities are
	forbidden. Even where this is the case, it does not necessarily follow
	that perturbation theory is convergent because Weinberg's theorem
	does not exclude the possibility of much slower divergences:
	for any fluctuation which is outside the horizon,
	these divergences scale with the number of e-folds of expansion since
	the time of horizon exit.

	Whether divergences are fast or slow, however, the interpretation is the
	same. When we expand an expectation value of some operators as a series
	in a loop-counting parameter or the slow-roll parameters,
	we are developing a series expansion based on the cut-off associated with
	the theory.
	The role of the
	cut-off is played by the time at which we wish to evaluate
	the expectation value---which is usually chosen to be at the end of
	inflation, or some similar time where we wish to use the expectation
	value as an initial condition for classical cosmological perturbation
	theory in the later universe.
	The behaviour of any expectation value as a function of
	this cut-off is merely
	its time dependence. The problem arises because
	truncating
	the series at any finite order gives the appearance of divergences.
	If we compute an answer which is superficially divergent in this way,
	then we must find some other method
	to compute the time dependence of the expectation value before
	growing secular terms take perturbation theory out of our control.

	This point of view leads to a technique of computation in which we
	can separate calculations into a quantum initial condition
	\cite{Seery:2005gb}, for
	which we need all the machinery of the so-called in--in
	(or \emph{Schwinger--Keldysh}) formalism, and a subsequent classical
	evolution for which to a good approximation we need only the
	classical, homogeneous evolution equations.
	The details of this approach have been developed by many
	authors \cite{Lyth:2005fi,Seery:2005gb,Lyth:2006qz,Seery:2008qj,
	Weinberg:2008nf,Weinberg:2008si}.
	There is a possible difficulty
	if we allow the inflating volume to become too large, because
	we may then encounter a source of \emph{quantum}
	divergences
	which could invalidate the use of classical evolution
	equations even after horizon exit \cite{Miao:2006pn,Prokopec:2007ak,
	Seery:2007wf,Bartolo:2007ti,
	Enqvist:2008kt,Cogollo:2008bi,Rodriguez:2008hy},
	but provided we work within some
	patch of spacetime not much larger than the size of the present horizon
	such effects are likely to be negligible.
	
	Most recent work on studying non-gaussianities from inflation has
	centred on the evolution subsequent to horizon exit
	\cite{Choi:2007su,Sasaki:2008uc,Naruko:2008sq,Byrnes:2008wi},
	whereas the quantum initial condition
	has received comparatively
	less attention
	\cite{Maldacena:2002vr,Alishahiha:2004eh,Seery:2005gb,Chen:2006nt}.
	There is a good reason for this imbalance:
	although there are known to be controlled examples where
	significant non-linearity can be generated outside the horizon
	\cite{Byrnes:2008wi,Byrnes:2008zy,Sasaki:2008uc},
	it is very hard (with canonical kinetic terms)
	to construct a controlled calculation
	in which a significant effect arises from the initial condition.
	Indeed, the most useful tool for extracting predictions from
	the underlying quantum field theory---that is, the slow-roll
	expansion---generally has the effect of forcing correlations
	to be very small. A second interpretation of the analysis given in
	this paper is an example in which the initial condition is modified,
	by including a coupling to high-energy virtual quanta which belong
	to the electromagnetic field.
	As can be expected, it will not be possible to control the calculation
	in the regime where this modification dominates the initial condition.
	However,
	we will be able to obtain a bound on the characteristics of
	the interaction among scalar and electromagnetic quanta which
	guarantees that the calculation is not taken beyond our control.
	
	Throughout this paper, we use units in which $\hbar = c = 1$
	and the reduced Planck mass
	is set equal to unity, giving $\Planck \equiv (8 \pi G)^{-1/2} = 1$,
	where $G$ is Newton's gravitational constant. The background space time
	is de Sitter space with flat spatial slices and metric
	\begin{equation}
		\d s^2 = - \d t^2 + a(t)^2 \d \vect{x} \cdot \d \vect{x} ,
	\end{equation}
	in the $(-,+,+,+)$ sign convention. Some formulae are more
	conveniently written in terms of conformal time, defined locally
	by the rule $\d t = a(t) \, \d \eta$ and given explicitly by the
	quadrature $\eta = \int_\infty^t \d t' / a(t')$.
	Spacetime indices are labelled
	with Latin indices $\{ a, b, c, \cdots \}$;
	purely spatial indices are labelled with indices $\{ i, j, k, \cdots \}$;
	and indices in the space of scalar fields are given Greek labels
	$\{ \alpha, \beta, \gamma, \cdots \}$.
	
	Purely spatial vectors such as
	$\vect{x}$ or $\vect{k}$ are written in bold face and an infix
	dot denotes index contraction with the flat background spatial metric,
	so that $\vect{x} \cdot \vect{k} \equiv \sum_i x_i k_i \equiv
	x_i k_i$, where the summation symbol will usually be omitted.
	Note that both indices are lowered.
	This convention is used to interpret exponentiation of any square spatial
	matrix $\gamma_{ij}$, giving the rule
	$\exp(\gamma)_{ij} \equiv \sum_{n=0}^{\infty} (\gamma^n)_{ij}/n!$.
	Repeated \emph{spacetime}
	indices in complementary raised and lowered positions
	are summed using the full spacetime metric $g_{ab}$ according to the
	Einstein convention, as usual; this convention also applies
	to raised and lowered spatial indices with the substitution of
	the full spatial metric in contractions.
	
	The model used as an example in this paper is Einstein gravity
	coupled to some collection of light scalar fields $\phi^\alpha$
	and a single $U(1)$ gauge field $A_a$. The $U(1)$ gauge field is taken
	to have a kinetic term of the form $\lambda(\phi) F^{ab} F_{ab}$, where
	$F_{ab} \equiv \partial_a A_b - \partial_b A_a$ is the gauge-invariant
	field strength. The coupling, $\lambda(\phi)$, is determined
	by the vacuum expectation values of some or all of the light scalars.
	This model is studied in \S\ref{sec:couplings}, where the interactions
	among the scalar, tensor and gauge field perturbations
	are derived.
	In \S\ref{sec:magnetogenesis} the magnetogenesis mechanism
	\cite{Turner:1987bw,Ratra:1991bn,Davis:1995mv,Davis:2000zp}
	is briefly
	reviewed, following an analysis by Bamba \& Sasaki.
	In \S\S\ref{sec:loop-spectrum}--\ref{sec:loop-bispectrum}
	I compute
	the leading loop correction---%
	for the spectrum and bispectrum---%
	which comes from scalar fluctuations
	mixing with virtual quanta of the gauge field.
	The paper concludes with a discussion in \S\ref{sec:conclude}.
	An auxiliary calculation of a simple pure scalar loop correction is
	given in \ref{appendix:loop} to aid comparison of the methods used
	in the present paper with those of other authors.
	
	\section{Scalar--magnetic couplings in the inflationary Lagrangian}
	\label{sec:couplings}
	
	Our starting point is Einstein gravity
	coupled to a collection of scalar fields $\phi^\alpha$,
	some of which are
	responsible for inflation, with the addition of a $U(1)$ gauge
	field $A_a$.
		
	\subsection{The gauge-fixed Einstein--scalar--vector action}
	As discussed in \S\ref{sec:introduction}, the normalization of the
	gauge field is taken to be controlled by some non-canonical coupling
	$\lambda(\phi)$ which is determined by the vacuum expectation values
	of some subset of the scalar fields. The action is
	\begin{equation}
		S = \frac{1}{2} \int \d^4 x \; \sqrt{-g} \left(
			R - \grad_a \phi^\alpha
			\grad^a \phi_\alpha - 2V - \frac{1}{2} \lambda(\phi)
			F^{ab} F_{ab} \right ),
		\label{eq:action}
	\end{equation}
	where the $U(1)$ field strength is defined by
	$F_{ab} \equiv \partial_a A_b - \partial_b A_a$ and
	$R$ is the spacetime Ricci scalar. The potential
	$V(\phi)$ is any reasonably smooth function
	which will generically depend on all the
	$\phi^{\alpha}$ and is arbitrary except that in order
	for the analysis which follows to apply it must support
	an epoch of inflation,
	at least for some range of values of the scalar fields.
	
	The background spacetime is taken to be homogeneous and isotropic,
	so the gauge field has no expectation value, up to configurations
	which are pure gauge. One can therefore assume that $A_a$ is
	a perturbation, which will generically couple to the scalar and
	gravitational degrees of freedom in Eq.~\eref{eq:action}. To study this
	system of coupled perturbations it is especially convenient to
	write the spacetime metric (including the effect of gravitational
	fluctuations) in the so-called Arnowitt--Deser--Misner (ADM) form,
	\begin{equation}
		\d s^2 = - N^2 \, \d t^2 + h_{ij}(\d x^i + N^i \, \d t)
			(\d x^j + N^j \, \d t) ,
	\end{equation}
	where $N$ (the ``lapse function'') and $N^i$ (the ``shift vector'')
	are not dynamical degrees of freedom,
	but instead are determined by constraint equations. The spatial
	metric $h_{ij}$ contains propagating tensor modes, and
	depending on the gauge it may also contain
	propagating scalar modes. After gauge-fixing
	$h_{ij}$, and solving the constraints, $N$ and $N^i$ can be written
	in terms of the propagating degrees of freedom in $h_{ij}$ and the
	$\phi^\alpha$. This is a great simplification in concrete
	calculations.
	
	Consider first the gauge sector. The ADM decomposition is based on
	a so-called ``3+1'' split of spacetime into three spatial dimensions
	and one timelike dimension. Many formulae are simplified by making
	an analogous decomposition of the vector potential,
	writing $A_a \equiv (\rho,\omega_i)$ where $\rho$ is the timelike
	component, and $\omega_i$ is a spatial 3-vector. It will transpire that
	$\rho$ is not a dynamical field, but is removed by a constraint
	associated with the gauge invariance of $A_a$. Once $\rho$ has been
	removed a further gauge fixing condition must be applied to $\omega_i$,
	which leaves the expected two physical polarizations of a massless
	gauge boson.
	
	When expressed in terms of $\rho$ and $\omega_i$, the part of the
	action involving the gauge field can be written
	\begin{equation}
		S \supseteq \frac{1}{2} \int \d^3 x \, \d t \; N\sqrt{h}
		\left\{ - \frac{\lambda}{2} h^{im} h^{jn} f_{ij} f_{mn} +
			\frac{\lambda}{N^2} h^{ij} \gdiff_i \omega \gdiff_j \omega ,
		\right\} ,
	\end{equation}
	where we have defined a useful quantity $\gdiff_i \omega$ by the rule
	\begin{equation}
		\gdiff_i \omega \equiv \dot{\omega}_i - \partial_i \rho
			+ f_{ij} N^j
	\end{equation}
	(denoting a time derivative with respect to $t$ by an overdot)
	and $f_{ij}$ is the \emph{spatially}
	gauge-invariant 3-vector field strength,
	$f_{ij} \equiv \partial_i \omega_j - \partial_j \omega_i$.
	The original gauge invariance associated with $A_a$ corresponding
	to $U(1)$ gauge transformations was $A_a \mapsto A_a + \partial_a
	\Lambda$ for some arbitrary spacetime-dependent function $\Lambda$.
	Under such a transformation, $\rho$ and $\omega_i$ undergo separate
	transformations determined by
	\begin{eqnarray}
		\rho & \mapsto \rho + \dot{\Lambda} , \\
		\omega_i & \mapsto \omega_i + \partial_i \Lambda .
	\end{eqnarray}
	It follows that $\gdiff_i \omega$ is gauge invariant, and therefore
	so is any action built out of $f_{ij}$ and $\gdiff_i \omega$ alone.
	
	The ADM action for $\rho$ and $\omega_i$ is singular, because
	the lagrangian is degenerate along pure gauge directions. It therefore
	cannot be quantized as it stands, but must be put into a form
	suitable for quantum mechanical calculations by adding a term of the
	form $s (c^\ast f)$, where $s$ is a so-called Slavnov operator,
	\begin{equation}
		s \equiv b (\delta \omega_i) \frac{\delta}{\delta \omega_i}
			- h \frac{\delta}{\delta c^\ast} ,
	\end{equation}
	$h$ is an auxiliary field, and $b$, $c^\ast$ are a pair of
	anti-commuting ghost fields which nevertheless obey Bose-Einstein
	statistics. The function $f$ is chosen to remove the degeneracy
	along gauge directions, but is otherwise essentially arbitrary provided
	that it is not invariant under gauge transformations. It is usually
	described as a gauge-fixing function.
	For the purposes of the present paper the most appropriate choice
	is an analogue of the Lorentz--Coulomb gauge, specified by
	$f = \partial_i \omega_i$.%
		\footnote{Recall that in the summation convention which is being
		used here, $\partial_i \omega_i \equiv \sum_i \partial_i
		\omega_i$.}
	It might have been thought that the covariant condition
	$f' = \partial^i \omega_i$ would be more appropriate in order to maintain
	manifest spatial covariance. Ultimately, however, we will be performing
	calculations in a version of the interaction picture in which it
	is simple to compute with $f$, but more complicated to compute with $f'$.
	It is immaterial whether we choose $f$ or $f'$, and since there are no
	serious consequences associated with losing manifest spatial covariance
	we will stick with $f$. (If desired, the spatially covariant
	action and Feynman rules can be obtained from those given here
	by the replacement $\delta^{ij} \mapsto h^{ij}$ in the gauge-fixing
	terms written in Eqs.~\eref{eq:gauge-fixed-action}--%
	\eref{eq:three-boson-interaction} below.)
	The total action is
	invariant under a quantized form of the original gauge symmetry,
	usually known as a BRST symmetry, which is generated by $s$.
	In the classical theory, the gauge condition is enforced by solving
	the constraint $f=0$ for one linear combination
	of the components of $\omega_i$. This removes one degree of freedom
	from the theory.
	
	After inserting the action into a path integral and
	appropriately integrating out the auxiliary field $h$, one finds that
	the ghosts $b$ and $c^\ast$ decouple and contribute only to an
	irrelevant overall normalization. The result can be written as a path
	integral over the action
	\begin{equation}
		S = \frac{1}{2} \int \d^3 x \, \d t \; \sqrt{h}
		\left( N B_1 + \frac{1}{N} B_{-1} \right) ,
		\label{eq:gauge-fixed-action}
	\end{equation}
	where the quantities $B_1$ and $B_{-1}$ are defined by
	\begin{eqnarray}
		B_1 & \equiv
			R - h^{ij} \partial_i \phi^\alpha \partial_j \phi_\alpha - 2V
			- \frac{\lambda}{2} h^{im} h^{jn} f_{ij} f_{mn}
			- \frac{1}{\xi} \delta^{ij} \delta^{mn} \partial_i \omega_j
			\partial_m \omega_n \\
		B_{-1} & \equiv
			E^{ij} E_{ij} - E^2 + \pi^\alpha \pi_\alpha +
			\lambda h^{ij} \gdiff_i \omega \gdiff_j \omega .
	\end{eqnarray}
	In these formulae, $R$ is the spatial Ricci curvature associated with
	$h_{ij}$; the quantities
	$\pi^\alpha \equiv \dot{\phi}^\alpha - N^j \partial_j \phi^\alpha$ are a
	collection of momenta associated with the scalar fields; and
	$E_{ij} = \dot{h}_{ij}/2 - \grad_{(i} N_{j)}$ is the gravitational
	momentum, where $\grad_i$ is the covariant derivative compatible with
	$h_{ij}$.
	
	\subsection{The constraint equations}
	The physical degrees of freedom in the action~\eref{eq:gauge-fixed-action}
	are a collection of propagating modes associated with the scalars,
	$\phi^\alpha$, together with modes arising from the components of
	$h_{ij}$. These are the fields whose time derivatives appear in the
	action.
	On the other hand, the
	fields $N$, $N^i$ and $\rho$ do not appear in the action with any
	time derivatives, and therefore are associated with constraints.
	These constraints can be satisfied and the unwanted degrees
	of freedom eliminated by simply solving for
	$N$, $N^i$ and $\rho$ in terms of the other fields in the system, and
	substituting the result back into the action.
	
	The constraints associated with $N$ and $N^i$ are well-known, and
	are modified here only by the presence of a component in the
	action associated with a gauge boson. The $N$ constraint is
	\begin{eqnarray}
		\fl\nonumber
		R - h^{ij} \partial_i \phi^\alpha \partial_j \phi_\alpha - 2V
		- \frac{\lambda}{2} h^{im} h^{jn} f_{ij} f_{mn}
		- \frac{1}{\xi} \delta^{ij} \delta^{mn} \partial_i \omega_j
		\partial_m \omega_n \\ \mbox{} -
		\frac{1}{N^2} \left( E_{ij} E^{ij} - E^2 + \pi^\alpha \pi_\alpha
		+ \lambda h^{ij} \gdiff_i \omega \gdiff_j \omega \right)
		= 0 ,
		\label{eq:lapse-constraint}
	\end{eqnarray}
	and the $N^i$ constraint is
	\begin{equation}
		\grad^i \left\{
			\frac{1}{N} \left( E_{ij} - h_{ij} E \right)
		\right\}
		= \frac{\pi^\alpha}{N} \partial_j \phi_\alpha -
		\frac{\lambda}{N} h^{ik} \gdiff_i \omega f_{kj} .
		\label{eq:shift-constraint}
	\end{equation}
	On the other hand, from integrating out $\rho$ we obtain
	a very simple constraint
	\begin{equation}
		\grad_j \left(
			\frac{\lambda}{N} h^{ij} \gdiff_i \omega
		\right) = 0 .
		\label{eq:rho-constraint}
	\end{equation}
	
	Consider first the $N^i$ constraint. We take the background
	field configuration to be spatially homogeneous and isotropic,
	so that the $\phi^\alpha$ are functions of $t$ only, with small
	perturbations $\delta\phi^\alpha$ which satisfy the smallness
	condition $|\delta\phi^\alpha| \ll |\phi^\alpha|$. If inflation has been
	ongoing for an exponentially large number of e-folds then this may
	be a poor approximation over the whole inflating volume, owing to
	back reaction effects which can cause large fluctuations in the
	scalar expectation values between widely separated regions. However, in
	any region of spacetime in the neighbourhood of exit from
	inflation this field theory is likely to be an accurate effective
	description.
	
	To parametrize the degrees of freedom associated with the spatial
	metric, we write $h_{ij} = a^2(t) (\e{\gamma})_{ij}$, where
	$\gamma_{ij}$ is a spatial $3 \times 3$ matrix whose components are
	taken to be of the same magnitude as $\delta\phi^\alpha$.
	One then aims for a perturbative solution in $\delta\phi^\alpha$
	and $\gamma_{ij}$, with the gauge field $\omega_i$ also taken to be
	perturbative and of the same formal magnitude. This implies that the
	gauge field does not modify the background evolution of
	the scalar fields. We define quantities
	$\alpha_n$, $\vartheta_n$ and $\beta_{nj}$ by writing
	\begin{eqnarray}
		N & = 1 + \sum_{n = 1}^{\infty} \alpha_n \\
		N_i & = \sum_{n = 1}^{\infty}
			\left( \partial_i \vartheta_n + \beta_{nj} \right) ,
	\end{eqnarray}
	where each of $\alpha_n$, $\vartheta_n$ and $\beta_{nj}$ is taken to
	contain exactly $n$ powers of the perturbations, and
	$\partial_j \beta_{nj} = 0$ for all $n$.
	
	At $\Or(1)$, the $N$ constraint~\eref{eq:lapse-constraint}
	gives the Friedmann equation for the background,
	\begin{equation}
		3 H^2 = \frac{1}{2} \dot{\phi}^\alpha \dot{\phi}_\alpha + V .
	\end{equation}
	At $\Or(\delta\phi)$, one obtains an equation for the scalar component
	of the shift vector, $\vartheta_1$,
	\begin{equation}
		\frac{4H}{a^2} \partial^2 \vartheta_1 = - 2 V_{,\alpha}
		\delta\phi^\alpha - 2 \dot{\phi}^\alpha \delta\dot{\phi}_\alpha +
		2\alpha_1 ( - 6H^2 + \dot{\phi}^\alpha \dot{\phi}_\alpha ) .
		\label{eq:vartheta1}
	\end{equation}
	To obtain equations for $\alpha_1$ and the vector component
	$\beta_{1j}$ one must return to the shift constraint,
	Eq.~\eref{eq:shift-constraint}. One finds $\beta_{1j} = 0$
	and
	\begin{equation}
		\alpha_1 = \frac{\dot{\phi}^\alpha \delta\phi_\alpha}{2H} .
		\label{eq:alpha1}
	\end{equation}
	For the purpose of computing interactions between the
	$\delta\phi^\alpha$, $\gamma_{ij}$ and $\omega_i$ to third order,
	it turns out to be unnecessary to obtain any terms in the lapse
	or shift which are of higher order in the perturbations
	than Eqs.~\eref{eq:vartheta1}--\eref{eq:alpha1}.
	Although such terms are present in a na\"{\i}ve expansion of the
	action, they are removed by the
	constraints~\eref{eq:lapse-constraint} and~\eref{eq:shift-constraint}
	\cite{Chen:2006nt}.
	Eqs.~\eref{eq:vartheta1}--\eref{eq:alpha1}, together with the
	constraint $\beta_{1j} = 0$, are exactly the solutions for the lapse
	and shift which were found by Maldacena in a theory with no gauge
	boson. Indeed, at first order
	(although not above), neither the gauge boson $\omega_i$ or
	any possible tensor modes $\gamma_{ij}$ contribute to $N$ or $N^i$.

	To solve the $\rho$ constraint one makes an analogous expansion of
	$\rho$ in powers of the perturbations, writing
	$\rho = \sum_{n = 1}^{\infty} \rho_n$
	with $\rho_n$ containing exactly $n$ powers of perturbations.
	At $\Or(\delta\phi)$ this gives
	\begin{equation}
		\partial^2 \rho_1 = \partial_i \dot{\omega}_i .
	\end{equation}
	
	\subsection{The Gaussian theory: Quadratic terms}
	We are now in a position to study the evolution of the fluctuations
	$\delta\phi^\alpha$, $\gamma_{ij}$ and $\omega_i$, together with their
	interactions. Any terms in the action which are linear in the
	perturbations vanish as a consequence of the background equations of
	motion. The leading non-trivial terms are therefore quadratic.
	Any theory defined by purely quadratic terms is free and gives rise to
	gaussian statistics. Therefore, the leading correction to gaussian
	statistics comes from interaction terms at cubic order or above.
	The details of these interactions
	can be calculated by treating them as small perturbations to the
	quadratic pieces, which we suppose still supply the dominant
	evolution. This formulation is equivalent to the interaction
	picture in the canonical formalism.
	
	At quadratic order, the perturbations $\delta\phi^\alpha$,
	$\gamma_{ij}$ and $\omega_i$ decouple and do not communicate with each
	other. The action therefore breaks into a sum of terms for each
	fluctuation which can be treated separately.
	
	Consider first the fluctuations in the light scalar fields
	$\delta\phi^\alpha$, which are described by the action
	\begin{eqnarray}
		\fl\nonumber
		S_2 \supseteq \frac{1}{2} \int \d^3 x \, \d t \; a^3 \Bigg\{
			\delta\dot{\phi}^\alpha \delta\dot{\phi}_\alpha -
			\frac{1}{a^2} \partial_i \delta \phi^\alpha
			\partial_i \delta \phi_\alpha - V_{\alpha\beta}
			\delta\phi^\alpha \delta\phi^\beta -
			\frac{2\dot{\phi}^\alpha}{a^2} \partial_j \vartheta_1
			\partial_j \delta\phi_\alpha \\
			\mbox{} + \alpha_1 \left[
				- \frac{4H}{a^2} \partial^2 \vartheta_1 -
				2 V_\alpha \delta\phi^\alpha - 2 \dot{\phi}^\alpha
				\delta \dot{\phi}_\alpha + \alpha_1 ( - 6H^2 +
				\dot{\phi}^\alpha \dot{\phi}_\alpha )
			\right]
		\Bigg\} .
	\end{eqnarray}
	In comparison the tensor modes
	$\gamma_{ij}$ obey a very simple action, containing only
	the minimal kinetic term
	\begin{equation}
		S_2 \supseteq \frac{1}{8} \int \d^3 x \, \d t \; a^3 \left\{
		\dot{\gamma}_{ij} \dot{\gamma}_{ij} - \frac{1}{a^2}
		\partial_k \gamma_{ij} \partial_k \gamma_{ij} \right\} .
	\end{equation}
	The action controlling the evolution of the gauge field satisfies
	\begin{equation}
		\fl
		S_2 \supseteq \frac{1}{2} \int \d^3 x \, \d t \; a \left\{
			\lambda ( \dot{\omega}_i - \partial_i \rho_1 )
			( \dot{\omega}_i - \partial_i \rho_1 )
			- \frac{\lambda}{2 a^2} f_{ij} f_{ij} - \frac{1}{\xi a^2}
			\partial_i \omega_i \partial_j \omega_j
		\right\} .
		\label{eq:gauge-quadratic}
	\end{equation}
	
	\subsection{Interactions: Cubic terms}
	The leading interactions among the $\gamma_{ij}$ and a single scalar
	degree of freedom were obtained by Maldacena, and are not affected
	by the presence of a gauge field. 
	However, in addition to the terms found by Maldacena there are now
	cubic interactions which involve gauge bosons together
	with the $\delta\phi^\alpha$ or $\gamma_{ij}$. Because the spacetime
	gauge connexion $A_a$ appeared in the original action
	Eq.~\eref{eq:action} quadratically, all these cubic interactions involve
	\emph{two} gauge bosons and only a single other field. There is
	therefore a term which describes the interaction of two gauge fields
	with a scalar,
	\begin{eqnarray}
		\fl\nonumber
		S_3 \supseteq \frac{1}{2} \int \d^3 x \, \d t \; a^3 \Bigg\{
			\frac{\lambda_\alpha \delta\phi^\alpha}{a^2}
			\left( \dot{\omega}_i \dot{\omega}_i - 2 \dot{\omega}_i
			\partial_i \rho_1 + \partial_i \rho_1 \partial_i \rho_1 \right)
			+ \frac{2 \lambda}{a^4} \dot{\omega}_i f_{ij} \partial_j
			\vartheta_1
			- \frac{\lambda_\alpha \delta\phi^\alpha}{2 a^4}
			f_{ij} f_{ij} \\ \mbox{} + \alpha_1 \left[
				- \frac{\lambda}{2a^4} f_{ij} f_{ij} -
				\frac{1}{\xi a^4} \partial_i \omega_i \partial_j \omega_j
				- \frac{\lambda}{a^2} \left(
					\dot{\omega}_i \dot{\omega}_i -
					2 \dot{\omega}_i \partial_i \rho_1 +
					\partial_i \rho_1 \partial_i \rho_1
				\right)
			\right]
		\Bigg\} .
		\label{eq:boson-scalar}
	\end{eqnarray}
	There is also an interaction between two gauge fields and a single
	$\gamma_{ij}$,
	\begin{equation}
		\fl
		S_3 \supseteq \frac{1}{2} \int \d^3 x \, \d t \; a^3 \Bigg\{
			- \frac{\lambda}{a^2} \gamma_{ij}
			\left( \dot{\omega}_i \dot{\omega}_j - 2 \dot{\omega}_i
				\partial_j \rho_1 + \partial_i \rho_1 \partial_j \rho_1
			\right)
			+ \frac{\lambda}{a^4} \gamma_{ij} f_{mi} f_{mj}
		\Bigg\}
	\end{equation}
	In principle there is also a self-interaction between three
	gauge bosons, described by the term
	\begin{equation}
		S_3 \supseteq \frac{1}{2} \int \d^3 x \, \d t \; a^3 \left\{
			- \frac{2 \lambda}{a^2} \dot{\omega}_i \partial_i \rho_2
		\right\} .
		\label{eq:three-boson-interaction}
	\end{equation}
	We will see in \S\ref{sec:magnetogenesis} below that it is possible
	to make a choice of gauge in which this interaction only involves
	the unphysical polarization of $\omega_i$. Therefore, in this gauge
	it decouples from all physical amplitudes, although it may remain
	present in more general gauges.
	
	\section{Magnetogenesis}
	\label{sec:magnetogenesis}
	
	In this section the process of magnetogenesis is briefly reviewed,
	with the aim of establishing the relevant
	notation and formulae which will be required for a calculation of
	loop corrections in
	\S\S\ref{sec:loop-spectrum}--\ref{sec:loop-bispectrum}.
	Our starting point is the assumption that the magnetic field
	arises from the coupling $\lambda F^{ab} F_{ab}$
	between the gauge field and the scalar sector. This calculation has
	been given in some generality by Bamba \& Sasaki
	\cite{Bamba:2006ga}, whose methods
	we follow. (See also Refs. \cite{Giovannini:2002ki,Giovannini:2003yn,
	Bertolami:2005np}.)
	
	When calculating processes involving exchange of virtual gauge bosons
	in Minkowski space it is usually a useful calculational check
	to leave the constant $\xi$ arbitrary. Indeed, since physical quantities
	do not depend on $\xi$ it must cancel out in any correct computation.
	Unfortunately, when attempting to compute the gauge field propagator
	in a time-dependent background, the presence of the $\xi$ term is an
	obstruction to solving the propagator equation. To simplify the
	process, it is helpful to take the limit $\xi \rightarrow 0$.
	For a gauge-fixing functional $f$ this directly enforces the constraint
	$f = 0$ in order that the action remain non-singular. Hence, instead
	of integrating over the \emph{three} components of $\omega_i$, we
	must make the decomposition
	\begin{equation}
		\omega_i(\vect{x},t) = \sum_{\sigma \in \pm}
		\int \frac{\d^3 k}{(2\pi)^3} e_i^\sigma(\vect{k})
		\omega_\sigma(k,t) \e{\im \vect{k} \cdot \vect{x}} ,
		\label{eq:gauge-decomposition}
	\end{equation}
	where the $\vect{e}^\sigma$ are so-called polarization vectors, labelled
	by a two-valued discrete index $\sigma$.
	The path integral should include only
	the \emph{two} physical polarizations $\omega_{\pm}$.
	The polarization vectors are chosen in such a way that
	$\vect{k} \cdot \vect{e}^\sigma(\vect{k}) = 0$
	for each $\sigma$, and are normalized so that
	\begin{eqnarray}
		\vect{e}^\sigma(\vect{k}) \cdot \vect{e}^{\sigma'}(\vect{k})^\ast
		& = \delta^{\sigma \sigma'} \\
		\sum_{\sigma \in \pm} e_i^\sigma(\vect{k}) e_j^\sigma(\vect{k})^\ast
		& = \delta_{ij} - \frac{k_i k_j}{k^2} .
		\label{eq:gauge-properties}
	\end{eqnarray}
	After carrying out this reduction, note that the apparent
	three-boson interaction~\eref{eq:three-boson-interaction}
	couples only to the unphysical degree of freedom in $\omega_i$.
	It can formally be removed after integrating by parts.
	It follows that this interaction
	is not physical.
	Note also that $\rho_1 = 0$ in this gauge, so all $\rho$ terms
	drop out of the action for $\omega_i$ up to cubic order.
	
	With these choices,
	the propagator for the gauge field is obtained by inverting
	the differential operator which appears under the integral in
	Eq.~\eref{eq:gauge-quadratic}.
	Suppose we write
	\begin{equation}
		\langle \omega_\sigma(\vect{k}_1,t) \omega_{\sigma'}(\vect{k}_2,t')
		\rangle = (2\pi) \delta(\vect{k}_1 + \vect{k}_2)
		\delta_{\sigma \sigma'} G_k(t,t') .
	\end{equation}
	The equation which determines $G_k$ is
	\begin{equation}
		\ddot{G}_k + \left( H + \frac{\dot{\lambda}}{\lambda} \right)
		\dot{G}_k + \frac{k^2}{a^2} G_k = - \frac{\im}{a \lambda}
		\delta(t-t') ,
		\label{eq:gauge-propagator}
	\end{equation}
	together with the boundary condition that
	in the limit $k/aH \rightarrow
	\infty$, where the fluctuation corresponding to this wavenumber
	cannot feel the curvature of spacetime,
	$G$ should approach the corresponding mode function from
	Minkowski space. This boundary condition plays an important role in
	the calculation. It corresponds to the stipulation that we begin with
	conventional Minkowski space quantum field theory in the ultra-violet,
	and then attempt to determine what this implies for fluctuations
	deep in the infra-red. Although it is possible to make more
	general choices of field theory in the ultra-violet, the assumption
	of flat space field theory is minimal and any admixture of different
	ultra-violet physics is usually subdominant to the Minkowski
	result. When we come to define what we mean by loop integrals in
	de Sitter space, which are also part of the specification of the
	ultra-violet behaviour of the theory, it will be necessary to take care
	that our definition does not destroy the property that we begin with flat
	space field theory at very high energies.
	
	Eq.~\eref{eq:gauge-propagator} cannot be solved exactly for a general
	choice of $\lambda(\phi)$. Instead, one can obtain a solution of
	Wentzel--Kramers--Brillouin (WKB)
	type which is valid inside the horizon, and can be
	matched onto a long wavelength solution which is valid outside the
	horizon. This is equivalent to using the flat space boundary
	condition deep inside the horizon to determine the size of fluctuations
	at horizon exit, and then using this as an initial condition
	for a classical superhorizon calculation.
	
	The solutions can be written most simply in terms of the conformal
	time coordinate $\eta$.
	In this variable, the relevant WKB solution is
	\begin{equation}
		G_k(\eta,\eta') = \frac{1}{2k}
		\frac{1}{\sqrt{\lambda(\eta)\lambda(\eta')}}
		\times \left\{
			\begin{array}{l@{\hspace{5mm}}l}
				\e{\im k (\eta - \eta')} & \eta < \eta' \\
				\e{\im k (\eta' - \eta)} & \eta' < \eta
			\end{array}
		\right. .
		\label{eq:wkb-propagator}
	\end{equation}
	It follows that the power spectrum of each polarization at horizon
	exit satisfies
	\begin{equation}
		\Ps_\ast = \frac{k^2}{4\pi^2} \frac{1}{\lambda_\ast} .
		\label{eq:magnetic-power}
	\end{equation}
	where `$\ast$' denotes evaluation at the time the mode with wavenumber
	$k$ exited the horizon. The $k$ dependence gives the spectrum a
	steep blue tilt, which means that if it remains unprocessed by
	new physics in the superhorizon regime it must have an essentially
	negligible magnitude on observable scales.
	
	It is sometimes said that a canonically normalized vector field does not
	receive a perturbation from inflation, based on the observation
	that positive and negative frequencies of the gauge field are not
	mixed as the universe expands
	\cite{Parker:1968mv,Parker:1969au}.
	It follows that if asymptotic \emph{in} and
	\emph{out} vacua can be defined there is no particle creation
	in the transition between the
	\emph{in}- and \emph{out}-vacuum. In the computation of
	inflationary observables, however, there is usually no natural
	\emph{out} region where a notion of particles make sense, nor any
	need to invoke such a region.
	Instead, we are interested primarily
	in whether expectation values of operators behave coherently over many
	e-folds of expansion, and in this sense a canonically normalized
	gauge field receives a fluctuation in exactly the same way as
	any light bosonic field.
	As we have already observed, however,
	the fluctuation which is imprinted in the precisely canonical case is
	very blue and entirely negligible on cosmologically relevant scales.
	One can think of this as a consequence of the fact that canonically
	coupled vector
	fluctuations must redshift like radiation, giving an extremely
	strong suppression for scales which exited the horizon
	early and have been redshifting for longer. Once modes re-enter
	the horizon they begin to oscillate and expectation values of their
	associated operators lose their coherence, which is consistent with
	Parker's observation that there is no asymptotic particle creation
	in this model \cite{Parker:1968mv}.%
		\footnote{I would like to thank David Lyth for helpful correspondence
		on this question.}
	
	The situation changes
	in the presence of the non-canonical coupling $\lambda(\phi)$.
	In this case
	Eq.~\eref{eq:magnetic-power} suggests that if $\lambda_\ast < 1$
	the power in fluctuations of the gauge field has been amplified
	in comparison with a vanilla model where $\lambda = 1$ for
	all time. However, this is
	misleading. Since the energy--momentum tensor associated with the gauge
	field is proportional to $\lambda$, when we compute the energy
	density stored in $\omega_i$ at horizon exit
	we obtain an answer which is independent of $\lambda_\ast$.
	It follows that there is a subtle distinction between this method of
	generating perturbations and the familiar case of scalar perturbations.
	In the simplest model of scalar perturbations, we generate
	fluctuations of the correct magnitude at horizon exit which are then
	conserved until horizon re-entry. In the case of $\omega_i$
	we do not make the physical fluctuations \emph{at horizon exit} any larger
	whether or not we include a coupling $\lambda$.
	It is evolution of $\lambda$
	\emph{after} horizon crossing which amplifies fluctuations in the
	gauge field and prevents them redshifting to zero;
	in this respect, the mechanism is similar to the curvaton
	example for purely scalar perturbations. We should therefore
	expect physical quantities to involve the ratio, $\lambda_2/\lambda_1$,
	of $\lambda$ at different times $\eta_1$ and $\eta_2$, which can be made
	large only if $\lambda$ evolves sufficiently strongly that there is
	a large hierarchy between its values at these times. In the limit
	of an observation made instantaneously at a moment in time,
	we can conclude that the physical effect must be proportional to
	$\lambda'/\lambda$, or a higher derivative, where $'$ denotes
	differentiation with respect to the conformal time.
	The power spectrum and bispectrum measured at horizon exit are
	examples of observations made instantaneously, and we shall see in
	\S\S\ref{sec:loop-spectrum}--\ref{sec:loop-bispectrum} below that
	they come proportional to powers of $\lambda'/\lambda$.
	An alternative means of breaking conformal invariance was
	considered in Ref.~\cite{Ashoorioon:2004rs}.
	
	Once a mode has left the horizon, its evolution is governed by
	Eq.~\eref{eq:gauge-propagator} in the limit $k/aH \rightarrow 0$.
	Discarding powers of gradients, the homogeneous solution for each
	polarization mode takes the form
	\begin{equation}
		\omega(\eta,\vect{x}) \equiv \omega_\ast(\vect{x}) +
		\lambda_\ast \omega_\ast'(\vect{x})
		\int_\ast^\eta \frac{\d \tau}
		{\lambda(\tau)} ,
		\label{eq:long-wavelength-mode}
	\end{equation}
	which depends on the value of $\omega$ and its derivative $\omega'$
	(where a prime $'$ denotes differentiation with respect to conformal
	time) on any initial hypersurface, $\eta = \eta_\ast$, provided that
	all relevant modes have left the horizon at that time. For a mode
	corresponding to a single wavenumber $k$ we can take this hypersurface
	to be the time of horizon exit, and the power spectra of $\omega$
	and $\omega'$ can be extracted from Eq.~\eref{eq:wkb-propagator}.
	If $\lambda$ is \emph{increasing} or decreasing less fast than
	$\eta$ then the integral in Eq.~\eref{eq:long-wavelength-mode} converges
	and each polarization is constant outside the horizon up to terms which
	decay exponentially fast in cosmic time.
	On the other hand, if $\lambda$
	decreases faster than $\eta$ then the integral diverges and
	each polarization grows rapidly. These possibilities correspond
	to a decreasing gauge coupling (or one which increases less fast
	than $a^{-1}$), or an increasing gauge coupling, respectively.
	
	At any time $\eta$ at which Eq.~\eref{eq:long-wavelength-mode}
	applies, the proper electric and magnetic energy densities
	were computed by Bamba \& Sasaki and are given by
	\cite{Bamba:2006ga,Bamba:2007sx}
	\begin{eqnarray}
		B_i & \equiv \frac{\sqrt{\lambda(\eta)}}{a(\eta)^2} \epsilon_{ijk}
		\partial_j \omega_k(\eta) =
		\frac{\sqrt{\lambda(\eta)}}{a(\eta)^2}
		(\grad \times \vect{\omega})_i ,
		\label{eq:magnetic-energy-density}
		\\
		E_i & \equiv \frac{\sqrt{\lambda(\eta)}}{a(\eta)^2} \omega_i' =
		\frac{\sqrt{\lambda(\eta)}}{a(\eta)}
		\frac{\d}{\d \eta} \vect{\omega}_i ,
		\label{eq:electric-energy-density}
	\end{eqnarray}
	where $\epsilon_{ijk}$ is the alternating tensor in three dimensions
	and the normalization has been chosen so that the electromagnetic
	energy density is given by $\rho_{\mathrm{EM}} =
	(\vect{B}^2 + \vect{E}^2)/2$, as usual. It follows from
	Eqs.~\eref{eq:magnetic-energy-density}--\eref{eq:electric-energy-density},
	Eq.~\eref{eq:long-wavelength-mode} and Eq.~\eref{eq:wkb-propagator}
	that the power spectrum of the proper magnetic energy density is
	\begin{equation}
		\Ps_B = \frac{1}{4\pi^2} \frac{\lambda}{\lambda_\ast}
		\left( \frac{k}{a} \right)^4
		\left|
			1 + \left( \im k - \frac{1}{2} \frac{\lambda_\ast'}{\lambda_\ast}
			\right) \int_\ast^\eta \d \tau \; \frac{\lambda_\ast}
			{\lambda(\tau)}
		\right|^2 ,
		\label{eq:magnetic-power-spectrum}
	\end{equation}
	and the power spectrum of the proper electric energy density is
	\begin{equation}
		\Ps_E = \frac{1}{4\pi^2} \frac{\lambda_\ast}{\lambda}
		\frac{k^2}{a^4} \left|
			\im k - \frac{1}{2} \frac{\lambda_\ast'}{\lambda_\ast}
		\right|^2 .
		\label{eq:electric-power-spectrum}
	\end{equation}
	Bamba \cite{Bamba:2007sx}
	(see also Giovannini \cite{Giovannini:2003yn}
	and Martin \& Yokoyama \cite{Martin:2007ue}) observed that
	if $\lambda$ is increasing (or decreasing less fast than $a^{-1}$),
	then the integral is dominated by its lower limit and one finds
	that $\Ps_E/\Ps_B \sim (\lambda_\ast/\lambda) \rightarrow 0$
	as $\eta \rightarrow 0$. On the other hand, if
	$\lambda$ is decreasing sufficiently fast to cause the integral
	to diverge then it is dominated by its upper limit, giving instead
	$\Ps_E/\Ps_B \sim (aH/k)$, which grows exponentially with the number
	of e-folds since horizon exit. One can conclude that in the first case
	one has a predominantly magnetic field at late times, with only an
	exponentially small admixture of electric field, whereas in the second
	case the situation is reversed.
	
	Let us focus on the case where a magnetic field is synthesized at late
	times, reserving the opposite case for the discussion
	in \S\ref{sec:conclude}.
	During radiation domination the electromagnetic field redshifts like
	the dominant constituent of the universe and therefore
	its relative density is conserved. If we assume prompt reheating
	after inflation, then
	the root mean square fluctuation in $\rho_B$ on a scale
	corresponding to comoving wavenumber $k$ has magnitude
	\begin{equation}
		|\vect{B}|_{\mathrm{rms}} \sim
		\left( \frac{k}{a} \right)^2
		\left( \frac{\lambda}{\lambda_\ast} \right)^{1/2}
	\end{equation}
	in Planck units. The proper wavenumber today on cluster scales
	is of order $k/a \sim \e{-140}$, and to seed a galactic dynamo it may
	be sufficient to produce fluctuations with magnitude \cite{Davis:1999bt}
	$|\vect{B}|_{\mathrm{rms}} \sim \e{-60}$ T.%
		\footnote{Our conventions for magnetic field strengths are
		measured in tesla, where $1 \, \mathrm{T} \sim \e{-120}$
		in Planck units.}
	Therefore to obtain a
	cosmologically interesting magnetic field, we require roughly
	$\lambda \sim \e{200} \lambda_\ast$
	\cite{Martin:2007ue}.
	Although the mechanism of magnetogenesis is quite insensitive to
	the dependence of $\lambda$ on the scalar fields which determine its
	value, a large class of models which invoke couplings of this form
	can be written in the ``dilaton-like'' form
	$\lambda(\phi) = \exp(\phi / M_\phi)$
	\cite{Bamba:2003av}, where $M_\phi$ is some
	characteristic mass scale. Such couplings have also been invoked
	in the context of models of dark energy
	\cite{Brax:2007hi}, where they may be subject to additional
	constraints \cite{Burrage:2007ew,Burrage:2008ii}.
	In order for a coupling of this form to
	yield the requisite hierarchy, $M_\phi$ must be chosen 
	to satisfy $M_\phi \approx \Delta\phi / 200$ where $\Delta\phi$ is
	the excursion of the field $\phi$ between horizon crossing
	and the end of inflation. If $\phi$ is a field driving a stage of
	chaotic inflation it can be supposed to move a distance
	perhaps of order $10$ in fundamental units, in which case
	\begin{equation}
		M_\phi \sim \frac{1}{20} .
		\label{eq:mass-scale}
	\end{equation}
	We will adopt this value of $M_{\phi}$ as the canonical one for
	magnetogenesis, although in practice $M_\phi$ will vary from model
	to model, and may be closer to the Planck scale.
	
	A variety of bounds are known on the energy density which can be present
	in magnetic fields at early times, either from direct detection
	\cite{Kahniashvili:2008hx} or indirect effects \cite{Caprini:2001nb}.
	These limits typically arise from constraints at the short
	wavelength end of the spectrum and imply that the extremely
	blue raw spectrum, Eq.~\eref{eq:magnetic-power}, must be processed
	into an approximately scale-invariant form. Exact scale invariance
	occurs for $M_\phi = \sqrt{2 \epsilon}/4$, which gives
	$M_{\phi}^{-1} \sim 10$,%
		\footnote{Compare, for example, with Eq.~(42) of 
		Ref.~\cite{Martin:2007ue}.}
	and it follows that for $M_{\phi}$ close to Eq.~\eref{eq:mass-scale}
	approximate scale invariance will apply.

	\section{Electric loop corrections to the scalar spectrum}
	\label{sec:loop-spectrum}
	
	In this section we return to the interaction between gauge bosons
	and the other fluctuation modes which are relevant during inflation.
	If we wish to make predictions for the anisotropy seen in the
	cosmic microwave background (CMB) then we are principally concerned
	with the power in scalar perturbations around the time of last
	scattering, because these dominate the density fluctuation in
	the primordial plasma. Such fluctuations are connected to observation
	by making
	predictions for the properties of the comoving curvature perturbation,
	$\R$, which on superhorizon scales is equivalent to the curvature
	perturbation on uniform density hypersurfaces, $\zeta$. In a model
	with many degrees of freedom there is a considerable simplification
	afforded by using $\zeta$, which can be computed using the
	so-called non-linear $\delta N$ formalism.
	
	As an initial condition, calculations using the $\delta N$ formalism
	require predictions for the correlations among the $\delta\phi^\alpha$
	around the time of horizon crossing. In this section we will
	compute a prediction for the two-point correlation of the
	$\delta\phi^\alpha$, taking into account the leading loop correction
	from exchange of virtual gauge bosons. This is
	presumably sufficient to make
	predictions for the power spectrum of the CMB,
	although it will transpire that there may be UV-divergent terms
	associated with the gauge transformation between $\zeta$ and
	the $\delta\phi^\alpha$ which are not captured by the $\delta N$
	formula. Therefore the final answer can be treated as an order
	of magnitude estimate only.
	
	In the following section we will compute the analogous correction
	for the three-point correlation function, which is necessary if we
	wish to study higher-order statistics.
	
	\subsection{Dominant contributions to the interaction vertex}
	Consider the vertex for interaction of two gauge bosons with a scalar
	particle, given by Eq.~\eref{eq:boson-scalar} in the limit $\xi
	\rightarrow 0$ with $\omega_i$ replaced by
	Eq.~\eref{eq:gauge-decomposition}. If we are only computing around
	the time of horizon crossing, then we can obtain a good
	approximation by truncating all quantities
	to leading order in the slow-roll expansion. In the interaction
	vertex, this leaves us with terms of the form
	\begin{equation}
		S_3 \supseteq \frac{1}{2} \int \d^3 x \, \d t \;
		a \; \lambda_\alpha \delta\phi^\alpha \left\{
			\dot{\omega}_i \dot{\omega}_i -
			\frac{1}{a^2}( \partial_i \omega_j \partial_i \omega_j -
			\partial_i \omega_j \partial_j \omega_i )
		\right\} .
		\label{eq:vertex-full}
	\end{equation}
	
	The first term in Eq.~\eref{eq:vertex-full} involves an interaction
	with $\dot{\omega}^2$, which according to
	Eq.~\eref{eq:electric-energy-density}
	can be thought of as a measure of the
	electric field intensity. The second term involves interactions with
	terms of the form $(\partial \omega)^2$, which
	according to Eq.~\eref{eq:magnetic-energy-density}
	are a measure of the magnetic field intensity. We can likewise
	imagine interactions which are dominated by the first or second term
	to represent interaction with the electric or magnetic field,
	respectively.
	This distinction is useful because if $\lambda$ is carrying a strong
	time dependence at horizon exit we expect the ambient electric field
	to be enhanced in comparison with the magnetic field, which depends only
	on spatial gradients.
	Indeed,
	this is true irrespective of whether $\lambda$ is increasing or
	decreasing, provided that it has a strong time dependence in either
	direction, because the properties of the fluctuations at horizon
	crossing do not determine whether the final configuration will be
	an electric or magnetic field. The initial condition involves a strong
	electric field in either case, but the final character of the field is
	only determined by the long-term evolution of $\lambda$ \emph{after}
	horizon crossing. 
	
	To see this in detail, it is simplest to use
	Eqs.~\eref{eq:gauge-decomposition} and~\eref{eq:wkb-propagator}
	to rewrite Eq.~\eref{eq:vertex-full} as an effective vertex
	which takes the form
	\begin{eqnarray}
		\fl\nonumber
		S_3 \supseteq \int \d \eta
		\int \frac{\d^3 k_1 \, \d^3 k_2 \, \d^3 k_3}{(2\pi)^9} \;
		(2\pi)^3 \delta(\vect{k}_1 + \vect{k}_2 + \vect{k}_3)
		\\ \mbox{} \times
		\frac{\lambda_\alpha}{2} \delta\phi^\alpha(\vect{k}_1,\eta)
		\alpha_{\pm \pm}^{ij} e^a_i(\vect{k}_2) e^b_j(\vect{k}_3)
		\omega_a(\vect{k}_2,\eta) \omega_b(\vect{k}_3,\eta) ,
		\label{eq:vertex}
	\end{eqnarray}
	where the vertex function
	$\alpha_{\iota_2 \iota_3}^{ij}$ is defined by
	\begin{equation}
		\alpha^{ij}_{\iota_2 \iota_3} =
		\delta_{ij} \left\{ \vect{k}_2 \cdot \vect{k}_3 +
			\left( \im \iota_2 k_2 + \frac{\Omega(\eta)}{\eta} \right)
			\left( \im \iota_3 k_3 + \frac{\Omega(\eta)}{\eta} \right)
		\right\} - k_{2j} k_{3i} .
		\label{eq:alpha-def}
	\end{equation}
	In this equation,
	$\Omega$ is an abbreviation for the dimensionless combination
	\begin{equation}
		\Omega \equiv \frac{1}{2} \frac{\lambda_\alpha}{\lambda}
		\frac{\dot{\phi}^\alpha}{H}
	\end{equation}
	and the $\pm$ symbols are chosen according the details of time
	ordering and the assignment of `$+$' or `$-$' vertices (to be described
	in \S\ref{sec:inin} below) among the gauge fields
	which participate in the vertex, and are fixed by the structure
	of the diagram in which Eq.~\eref{eq:vertex} is inserted.
	(In Eq.~\eref{eq:alpha-def}, we have temporarily abandoned our
	summation convention---for this equation only---in
	the interests of notational clarity:
	this choice of $\alpha^{ij}_{\iota_2 \iota_3}$ should be inserted
	directly in Eq.~\eref{eq:vertex} without concern for the position
	of the spatial indices $i$ and $j$.)
	The parameter $\Omega$ measures the hierarchy between the wavefunction
	of the gauge field and its time derivative, and arises
	from the interaction with the electric field.
	In the limit $|\Omega| \gg |k\eta|$, the total interaction is dominated
	by this electric piece and the details of the time ordering become
	irrelevant. This is true whenever the rate of change of the coupling
	$\lambda$ with the fields is tuned to satisfy
	\begin{equation}
		\frac{1}{2}
		\left( \frac{\dot{\phi}^\alpha}{H} \right)_\ast
		\left( \frac{\lambda_\alpha}{\lambda} \right)_\ast
		\gg
		\e{-N_\ast}
		\label{eq:electric-dominance} .
	\end{equation}

	In Eq.~\eref{eq:electric-dominance},
	`$\ast$' denotes evaluation at the time when the mode with wavenumber
	$k$ exited the horizon, or more precisely a small but non-zero
	number of e-folds $N_\ast \sim 1$ afterwards. This time should be
	chosen so that the fluctuations in scalar modes are close to their
	asymptotic superhorizon values, and the canonical commutation
	relation $[\delta\dot{\phi},\delta\phi] \sim \e{-N_\ast}$
	allows the $\delta\phi^\alpha$ to be treated as approximately
	classical quantities, but it should not be so late after
	horizon exit that appreciable evolution might have occurred, which
	would spoil the accuracy of the slow-roll approximation.
	Whenever Eq.~\eref{eq:electric-dominance} applies we can ignore
	the purely magnetic part of the interaction.
	It is important to note,
	however, that the electric interaction can only become
	dominant if at least one of the scalars which couple to
	$\omega_i$ is rolling on cosmological timescales, so that
	$\dot{\phi}^\alpha/H$ is not totally negligible for some
	$\phi^\alpha$. This is not really a restriction if one wishes to
	use this interaction for the purposes of magnetogenesis, because one
	is then dependent on $\lambda$ developing a large hierarchy between
	its value at horizon exit and its value at some much later time,
	such as the end of inflation, and this can occur only if some of the
	scalar fields are in motion.
	
	Alternatively, the scalar fields whose vacuum expectation values determine
	the magnitude of $\lambda$ may not be rolling during inflation, or
	the dependence of $\lambda$ on these fields may be so weak that
	$(\ln \lambda)_{\alpha}$ is never large enough for
	Eq.~\eref{eq:electric-dominance} to apply. In this limit the interaction
	is still interesting, but it is magnetically dominated and the
	momentum integral which describes the loop is somewhat
	more complicated to compute. For the remainder of this paper, we
	focus on the electric part of the interaction only.
	
	\subsection{The in--in formalism}
	\label{sec:inin}
	The appropriate formalism in which to compute expectation values
	of cosmological fluctuations is the so-called in--in formalism
	introduced by Schwinger. In the cosmological case we wish to
	compute expectation values of the form $\bra{0} O \ket{0}$
	for some collection of operators $O$, in the state $\ket{0}$ which
	following the discussion of ultra-violet physics
	given in \S\ref{sec:magnetogenesis}
	should be taken to be the Minkowski vacuum deep inside the horizon.
	We know from scattering calculations in Minkowski space that
	one can compute
	$\bra{\mbox{out}} O \ket{\mbox{in}}$ using a conventional Feynman
	path integral for any `in-state' $\ket{\mbox{in}}$ and
	`out-state' $\ket{\mbox{out}}$. It follows that
	after integrating over all possible
	out-states we can compute $\bra{\mbox{in}} O \ket{\mbox{in}}$
	using \emph{two} path integrals, which gives the so-called
	Schwinger--Keldysh path integral formula
	\begin{equation}
		\bra{\mbox{in}} O \ket{\mbox{in}} =
		\int [ \d \phi_+ \, \d \phi_- ]
		\; O \, \exp \left(
			\im S[\phi_+] - \im S[\phi_-]
		\right) ,
		\label{eq:inin}
	\end{equation}
	where the $\phi$ label the elementary fields in the theory and $O$
	is taken to be built out of either `$+$' or `$-$' fields but not both.
	The integral is defined by prescribing that only those $+$ and $-$ field
	configurations which begin in the appropriate vacuum
	$\ket{0}$ and end with coincident values at some late time
	are to be included in the integration. The precise choice of this
	late time is immaterial, provided it is chosen to be later than the
	time of evaluation of any field which appears in $O$.
	We will conventionally choose $O$ to be composed only from $+$ fields
	and carry the integral from past infinity to the time of observation,
	$\eta_\ast$.
	
	When coupling to gravity is taken into account, there can be a
	subtlety in the construction of the path integrals which appear in
	Eq.~\eref{eq:inin}. From a given formula for the lagrangian, one would
	ordinarily obtain the canonical momenta and construct the
	hamiltonian. Integrating over the coordinates and canonical momenta,
	with time evolution supplied by the hamiltonian, one arrives at the
	standard path integral. If the lagrangian is not quadratic in the
	momenta, however, then one cannot explicitly integrate them out
	\cite{Seery:2007we}.
	Instead, one must include their degrees of freedom in the
	path integral as ``derivative
	ghosts'' which compensate for the fact that the fields are not
	canonically normalized. They are associated with interactions which
	are cubic or higher in derivatives such as $\delta\dot{\phi}$.
	In our example, the only such interactions are associated with
	scalar fluctuations and for this reason we ignore derivative ghosts.
	Instead, they
	should instead included with scalar loop corrections, which
	presumably do not lead to large effects
	\cite{Seery:2007we,Dimastrogiovanni:2008af,Adshead:2008gk}.
	
\begin{figure}
		\begin{center}
			\hfill
			\begin{fmfgraph*}(60,60)
				\fmfpen{0.8thin}
				\fmfleft{l}
				\fmfright{r}
				\fmf{plain,tension=2,label=$\scriptsize\vect{k}_1$,
				     label.side=left}{l,v1}
				\fmf{plain,tension=2,label=$\scriptsize\vect{k}_2$,
				     label.side=right}{r,v2}
				\fmf{boson,left}{v1,v2}
				\fmf{boson,left}{v2,v1}
				\fmfv{label=$\scriptsize +$,label.angle=-125}{v1}
				\fmfv{label=$\scriptsize +$,label.angle=-55}{v2}
			\end{fmfgraph*}
			\hfill
			\begin{fmfgraph*}(60,60)
				\fmfpen{0.8thin}
				\fmfleft{l}
				\fmfright{r}
				\fmf{plain,tension=2,label=$\scriptsize\vect{k}_1$,
				     label.side=left}{l,v1}
				\fmf{plain,tension=2,label=$\scriptsize\vect{k}_2$,
				     label.side=right}{r,v2}
				\fmf{boson,left}{v1,v2}
				\fmf{boson,left}{v2,v1}
				\fmfv{label=$\scriptsize +$,label.angle=-125}{v1}
				\fmfv{label=$\scriptsize -$,label.angle=-55}{v2}
			\end{fmfgraph*}
			\hfill
			\begin{fmfgraph*}(60,60)
				\fmfpen{0.8thin}
				\fmfleft{l}
				\fmfright{r}
				\fmf{plain,tension=2,label=$\scriptsize\vect{k}_1$,
				     label.side=left}{l,v1}
				\fmf{plain,tension=2,label=$\scriptsize\vect{k}_2$,
				     label.side=right}{r,v2}
				\fmf{boson,left}{v1,v2}
				\fmf{boson,left}{v2,v1}
				\fmfv{label=$\scriptsize -$,label.angle=-125}{v1}
				\fmfv{label=$\scriptsize +$,label.angle=-55}{v2}
			\end{fmfgraph*}
			\hfill
			\begin{fmfgraph*}(60,60)
				\fmfpen{0.8thin}
				\fmfleft{l}
				\fmfright{r}
				\fmf{plain,tension=2,label=$\scriptsize\vect{k}_1$,
				     label.side=left}{l,v1}
				\fmf{plain,tension=2,label=$\scriptsize\vect{k}_2$,
				     label.side=right}{r,v2}
				\fmf{boson,left}{v1,v2}
				\fmf{boson,left}{v2,v1}
				\fmfv{label=$\scriptsize -$,label.angle=-125}{v1}
				\fmfv{label=$\scriptsize -$,label.angle=-55}{v2}
			\end{fmfgraph*}
			\hfill
			\mbox{}
		\end{center}
		\caption{\label{fig:twopoint}Schwinger-formalism diagrams for the
		loop correction to the scalar two-point function which arise from
		mixing with gauge bosons. Straight lines indicate scalar quanta,
		which only appear on the external legs. The interior loop,
		composed of wavy lines, indicates mixing with virtual quanta
		borrowed from the ambient electric and magnetic fields. The
		fields associated with external legs are always of $+$ type,
		whereas the vertices can be of $+$ or $-$ type. The $(+,+)$
		and $(-,-)$ diagrams form one complex conjugate pair, and the
		$(+,-)$ and $(-,+)$ diagrams form another.}
	\end{figure}
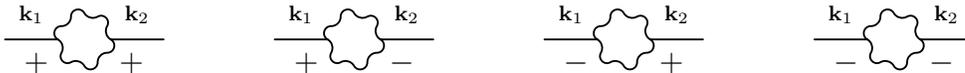

	The doubling of degrees of freedom in Eq.~\eref{eq:inin} implies that
	when we expand the path integral into diagrams we encounter extra
	graphs, which mix vertices constructed from $+$ and $-$ fields.
	These vertices collectively ensure that all expectation values are real.
	If we apply Eq.~\eref{eq:inin} to the calculation of the scalar
	two-point function $\langle \delta\phi^\alpha(\vect{k}_1)
	\delta\phi^\beta(\vect{k}_2) \rangle_\ast$, one obtains the diagrams
	shown in Fig.~\ref{fig:twopoint}.
	
	The diagrams in Fig.~\ref{fig:twopoint} divide
	into two pairs of complex conjugates, corresponding to the
	$(+,+)$ and $(-,-)$ diagrams in one pair and the $(+,-)$ and
	$(-,+)$ diagrams in another. Consider first the $(+,+)$ diagram.
	This makes a contribution to the scalar two-point function
	of the form%
		\footnote{In writing these and all subsequent expressions,
		I have chosen to nest the integrals which result from
		mixed $(+,-)$ and $(-,+)$-type diagrams. Alternatively,
		it is possible to factorize these contributions, obtaining
		expressions which are manifestly non-singular for all
		momenta $k_i$ \cite{Seery:2008ax,Adshead:2009cb,Chen:2009bc}.
		Whichever method is chosen, the answer is always the same.
		I would like to thank
		Peter Adshead, Richard Easther, Eugene Lim,
		Martin Sloth and Filippo Vernizzi for correspondence on this
		issue. }
	\begin{eqnarray}
		\fl\nonumber
		\langle \delta\phi^\alpha(\vect{k}_1) \delta\phi^\beta(\vect{k}_2)
		\rangle_\ast \supseteq
		\frac{H_\ast^4}{32 \prod_i k_i^3}
		\left( \frac{\lambda_\alpha \lambda_\beta}{\lambda^2} \right)_\ast
		\Omega_\ast^4
		(1 + \im k_1 \eta_\ast)(1 + \im k_2 \eta_\ast)
		\e{-\im \eta_\ast (k_1 + k_2)}
		\\ \nonumber \mbox{} \times
		\int \frac{\d^3 q \, \d^3 r}{q^3 r^3} \;
		P_2(\vect{q},\vect{r})
		\delta(-\vect{k}_1 - \vect{r} + \vect{q})
		\delta(-\vect{k}_2 - \vect{q} + \vect{r})
		\\ \mbox{} \times
		\left(
			\twoEi{k_1}{k_2}{k_1 + r + q}{k_2 - r - q} +
			\twoEi{k_2}{k_1}{k_2 + r + q}{k_1 - r - q}
		\right) ,
		\label{eq:twopoint-plusplus}
	\end{eqnarray}
	where $i \in \{ 1, 2 \}$,
	which depends on a four-parameter integral, defined by
	\begin{equation}
		\twoEi{\alpha_1}{\alpha_2}{\beta_1}{\beta_2} \equiv -
		\int_{-\infty}^{\eta_\ast} \frac{\d \tau}{\tau^2}
		\int_{-\infty}^{\tau} \frac{\d \eta}{\eta^2}
		(1 - \im \alpha_1 \eta) (1 - \im \alpha_2 \tau)
		\e{\im \beta_1 \eta} \e{\im \beta_2 \tau} .
		\label{eq:time-two}
	\end{equation}
	To make sense of this for real $\beta_1$ and $\beta_2$, the contours
	of integration for $\eta$ and $\tau$ must be deformed so that
	$\e{\im \beta_1 \eta}$ and $\e{\im \beta_2 \tau}$ are decaying
	for large $|\eta|$ and $|\tau|$, respectively.
	Indeed, this contour prescription
	follows automatically from the choice of vacuum $\ket{0}$ in
	Eq.~\eref{eq:inin}.
	In addition, $P_2(\vect{q},\vect{r})$ is a polynomial in the
	vector momenta which depends on the transverse structure
	of the gauge field propagator and the detailed momentum
	dependence at the vertex. It satisfies
	\begin{equation}
		P_2(\vect{q},\vect{r}) \equiv q^2 r^2 + (\vect{q}\cdot\vect{r})^2 .
		\label{eq:spectrum-polynomial}
	\end{equation}
	Eq.~\eref{eq:twopoint-plusplus} has been written in a form which
	emphasizes the symmetry between $\vect{k}_1$ and $\vect{k}_2$. In order
	to simplify it further, one can integrate out either $\vect{q}$ or
	$\vect{r}$, which leaves behind a single momentum-conservation
	delta-function, $\delta(\vect{k}_1 + \vect{k}_2)$, together with
	an integral over the remaining momentum. Since local field theories
	usually exhibit bad behaviour at high energies, we may expect
	this integral to receive a significant contribution from the region where
	the 3-momentum which circulates in the loop becomes large.
	Indeed, neglecting any powers of momentum which are introduced by
	the time integrals in Eq.~\eref{eq:time-two},
	it easy to see that the loop
	diverges at least as fast as $\int \d q$ in the ultra-violet, whereas
	it converges at least as fast as $\int q \, \d q$ in the infra-red.
	The time integral has dimension $q^2$, so it can only make this
	divergence worse at high energy, and improve convergence at low
	energy. To make sense of such an integral it
	must first be regularized, removing
	the contribution of arbitrarily
	high-energy modes,
	and made finite by applying a renormalization prescription.
	Once this is done, the ultra-violet sensitivity of the
	original integral implies that we can expect the loop to be dominated
	by contributions near some UV scale. In what follows, we use this
	to ignore effects associated with the infra-red cutoff.
	
	How are we to choose a cut-off in an expanding, inflationary spacetime?
	In flat space quantum field theory we are familiar with the use of a
	variety of
	regulators. The method of the renormalization group tells us that
	these are all equivalent in the continuum limit, after the introduction
	of counter-terms and a renormalization prescription.
	The simplest choice is a hard cutoff, $M$, on the invariant momenta
	associated with particles which circulate within the loops.
	To apply this cutoff in practice one must Wick rotate
	loop integrals to Euclidean spacetime where $M$ becomes
	a $SO(4)$-invariant cut-off on the Euclidean momentum.
	When we revert to Lorentzian signature, this procedure gives a
	$SO(3,1)$-invariant result and therefore preserves
	Lorentz invariance.
	As a result,
	the hard momentum cutoff
	in Minkowski space provides a popular means of estimating
	the magnitude of loops in an effective theory. The problem at hand
	is to find a way to perform similar estimates in de Sitter space.
	
	In any curved spacetime we do not have global Lorentz invariance,
	although according to the equivalence principle we must recover
	approximate local $SO(3,1)$ invariance in a small neighbourhood
	of any point.
	On the basis of the discussion in \S\ref{sec:magnetogenesis}
	it follows that this
	restoration of Lorentz invariance in the ultra-violet corresponds
	to choosing conventional flat space quantum field theory deep
	inside the horizon. Whatever regulator we pick, it is necessary to be
	sure that it does not conflict with local Lorentz invariance in the
	ultra-violet, or we shall obtain nonsensical results.
	Recently, van der Meulen and Smit \cite{vanderMeulen:2007ah}
	have pointed out that that a momentum cut-off at some characteristic
	scale $M$ should be taken to apply to \emph{proper} momenta, rather
	than comoving momenta, since only the former have physical significance.
	However, such a cut-off
	violates local Lorentz invariance for any finite value of the
	cut-off.%
		\footnote{I would like to thank Andrew Tolley for
		emphasizing this property of the proper momentum cut-off.
		A similar observation has been made in several places
		in the literature; see especially Refs.
		\cite{Niemeyer:2000eh,Kempf:2001fa,Ashoorioon:2004vm}.}
	This is because the cut-off grows between adjacent spatial slices
	which are taken at later and later times.
	If we pick some particular point in spacetime and ask whether
	Lorentz invariance is restored in a local neighbourhood of that
	point we find that no matter how small a patch of spacetime we choose,
	it necessarily intersects a sheaf of spatial slices.
	The growth in the cut-off transverse to the sheaf picks out a preferred
	direction, and the properties of the theory do not become invariant under
	local spacetime rotations within the patch. It follows that Lorentz
	invariance is broken.
	
	Breakdown of Lorentz invariance in the ultra-violet
	would have dramatic consequences for the stability of the theory.
	As time increases, the cut-off grows. This means that between two
	adjacent spatial slices, one taken at a slightly later time than the
	other, new quanta are introduced to our universe
	\cite{Brandenberger:1999sw}.
	These new quanta destabilize the theory, because although they are
	at very high energies when they are added to the description, they
	can scatter with each other to produce other quanta which are much
	softer. These scattered soft quanta, whose correlations are essentially
	random, impinge on superhorizon correlation functions and scramble their
	values. Thus, in the presence of the Lorentz-violating cut-off,
	correlation functions do not retain coherent values outside the horizon
	but instead behave as if they are coupled to a quantum noise bath
	and are swamped exponentially quickly by quantum noise.
	In this cut-off theory
	there is no quantum-to-classical transition, and presumably there
	are no conserved quantities outside the horizon.
	On the other hand, if we find a way to
	preserve Lorentz invariance then none
	of these undesirable features become manifest.
	One can think of this stabilizing property of Lorentz invariance
	as analogous to the conservation of the velocity of the centre of
	energy of any isolated system. This is a consequence of invariance
	under Lorentz boosts and implies, for example, that quantum effects
	associated with virtual ultra-violet quanta do not disturb the motion
	of isolated particles. A form of this analogy was earlier used by Lyth
	\cite{Lyth:1984gv}.
		
	Let us return to Eqs.~\eref{eq:twopoint-plusplus}--\eref{eq:time-two}
	and study the
	time and loop integrals in more detail. Along the way, we will
	encounter the features described in the previous paragraph.
	In~\ref{appendix:loop}
	an analogous analysis is given for the scalar loop which arises
	from a $V'''$ interaction, which has previously appeared in
	the literature \cite{vanderMeulen:2007ah,Bartolo:2007ti}.
	The $(+,+)$ diagram described by Eq.~\eref{eq:twopoint-plusplus}
	must be supplemented by a $(+,-)$ diagram, which gives a contribution
	corresponding to
	\begin{eqnarray}
		\fl\nonumber
		\langle \delta\phi^\alpha(\vect{k}_1) \delta\phi^\beta(\vect{k}_2)
		\rangle_\ast \supseteq
		- \frac{H_\ast^4}{32 \prod_i k_i^3}
		\left( \frac{\lambda_\alpha \lambda_\beta}{\lambda^2} \right)_\ast
		\Omega_\ast^4
		(1 + \im k_1 \eta_\ast)(1 - \im k_2 \eta_\ast)
		\e{- \im \eta_\ast (k_1 - k_2)}
		\\ \nonumber \mbox{} \times
		\int \frac{\d^3 q \, \d^3 r}{q^3 r^3}
		P_2(\vect{q},\vect{r})
		\delta(-\vect{k}_1 - \vect{r} + \vect{q})
		\delta(-\vect{k}_2 - \vect{q} + \vect{r})
		\\ \mbox{} \times
		\left(
			\twoEi{k_1}{-k_2}{k_1+r+q}{-k_2-r-q} +
			\twoEi{-k_2}{k_1}{-k_2-r-q}{k_1+r+q}
		\right) .
		\label{eq:twopoint-plusminus}
	\end{eqnarray}
	Once we have aggregated the contribution of
	Eq.~\eref{eq:twopoint-plusplus} and Eq.~\eref{eq:twopoint-plusminus}
	we must remember to add in their complex conjugates in order to
	account for the $(-,-)$ and $(-,+)$ diagrams which have not been
	written explicitly.
	
	After integration by parts in $\eta$ and $\tau$,
	the fundamental time integral given in Eq.~\eref{eq:time-two}
	can be re-expressed in the form
	\begin{eqnarray}
		\fl\nonumber
		\twoEi{\alpha_1}{\alpha_2}{\beta_1}{\beta_2} =
		\left[ - \frac{1}{2 \eta_\ast^2} + \frac{\im}{\eta_\ast}
			\left( \alpha_2 - \alpha_1 + \frac{\beta_1 - \beta_2}{2}
			\right)
		\right]
		\e{\im (\beta_1 + \beta_2) \eta_\ast}
		+ \frac{\im}{\eta_\ast} (\beta_1 - \alpha_1)
			\e{\im \beta_2 \eta_\ast}
			\int_{-\infty}^{\eta_\ast} \frac{\d \tau}{\tau}
			\e{\im \beta_1 \tau}
		\\ \nonumber \mbox{}
		+ (\beta_1 + \beta_2)\left(\alpha_2 - \alpha_1 +
			\frac{\beta_1 - \beta_2}{2}\right)
			\int_{-\infty}^{\eta_\ast} \frac{\d \tau}{\tau}
			\e{\im (\beta_1 + \beta_2) \tau}
		\\ \mbox{}
		+ (\beta_1 - \alpha_1)(\beta_2 - \alpha_2)
			\int_{-\infty}^{\eta_\ast} \frac{\d \tau}{\tau}
			\e{\im \beta_2 \tau}
			\int_{-\infty}^{\tau} \frac{\d \eta}{\eta}
			\e{\im \beta_1 \eta} .
		\label{eq:time-two-asymptotic}
	\end{eqnarray}
	This equation shows primitive fast divergences but some of these will
	cancel and others are purely imaginary, which implies that they
	disappear when the contribution of all four diagrams is accounted for.
	Write $k_1 = k_2 = k$ and consider the deep ultra-violet region,
	where $q \approx r \gg k$. In this region, the $\beta_i$ terms
	are individually very large, approximately satisfying
	$\beta_i \sim \pm 2q \gg k$, so that the integrals are almost all
	very small---except for the integral with $\beta_1 + \beta_2$ in
	the exponent, since this combination is independent of $q$.%
		\footnote{Conversely, these integrals diverge like powers of
		$\ln|q \eta_\ast|$ near the infra-red cutoff, where $q$
		is a measure of the 3-momentum circulating in the loop.
		However, in this region such logarithms are suppressed by positive
		powers of $q$ and therefore the loop integral will be well-behaved.
		Indeed, for the purposes of the present calculation, we are assuming
		that the loop integral is dominated by exchange of virtual
		quanta near the ultra-violet cutoff, so these integrals
		(and other similar integrals to be encountered
		in \S\ref{sec:loop-bispectrum} while studying the loop-corrected
		three-point function) actually play no role in the analysis.}
	Collecting all necessary terms, expanding for small $|k\eta_\ast|$
	and keeping only contributions which are relevant in this limit,
	the expectation value can be written in the form
	\begin{equation}
		\langle \delta\phi^\alpha(\vect{k}_1)
		\delta\phi^\beta(\vect{k}_2)
		\rangle_\ast =
		(2\pi)^3 \delta(\sum_i \vect{k}_i)
		\frac{H^4_\ast}{32 \prod_i k_i}
		\left(
			\frac{\lambda^\alpha \lambda^\beta}{\lambda^2}
		\right)_\ast
		\Omega_\ast^4 I_2 ,
	\end{equation}
	with the left-over time and momentum dependence consolidated into
	an integral of the form
	\begin{equation}
		\fl
		I_2 = 4 \int \frac{\d^3 q}{q^3 |\vect{k} - \vect{q}|^3} \;
		P_2(\vect{q},\vect{k}-\vect{q})
		\Big\{
			k^2 - k (q + |\vect{k}-\vect{q}|)(N_\ast + \ln 2 -
			\EulerMascheroni - 1)
		\Big\} ,
		\label{eq:spectrum-momentum-integral}
	\end{equation}
	where $\EulerMascheroni \approx 0.57722$ is the Euler--Mascheroni
	constant and $N_\ast$ measures by how many e-folds the mode with
	wavenumber $k$ is outside the horizon at time $\eta_\ast$.
	
	Eq.~\eref{eq:spectrum-momentum-integral} is quadratically divergent.
	Introducing a cut-off $\Lambda$ (as yet unspecified)
	on the momentum circulating in the loop, this integral takes the
	form
	\begin{equation}
		\fl
		I_2 \approx - \frac{8 k}{\pi^2} \Lambda^2
		( N_\ast + \ln 2 + \EulerMascheroni - 1)
		+ \frac{4k^2}{\pi^2} \Lambda
		+ \frac{16 k^3}{3 \pi^2} \ln \frac{\Lambda}{\mu}
		( N_\ast + \ln 2 + \EulerMascheroni - 1) + \cdots ,
		\label{eq:spectrum-divergences}
	\end{equation}
	where `$\cdots$' denotes terms which are subdominant in the limit
	$\Lambda \rightarrow \infty$, and
	$\mu$ is an arbitrary scale, of the same dimensions as
	$\Lambda$, which has been introduced to make sense of the logarithm.
	It is clear that if we take $\Lambda$ to correspond to a proper
	momentum cut-off, which would take the form
	$\Lambda = a_\ast \tilde{\Lambda}$ for some constant $\tilde{\Lambda}$,
	then we introduce fast divergences as $a_\ast \rightarrow \infty$.
	These divergences correspond to hard quanta which are redshifted into the
	effective theory as the universe expands but subsequently scatter
	to produce soft quanta and contaminate the spectrum, as
	described above Eq.~\eref{eq:twopoint-plusminus}. Indeed, since the
	accumulation of such quanta is presumably highly incoherent, it
	seems unlikely that one can ascribe any definite value to the
	spectrum after horizon crossing.
	
	The key lesson I wish to draw from this example is that such
	divergent effects are fictional. If we begin with
	a Lorentz invariant theory valid at high scales and integrate out
	modes above a proper momentum cut-off,
	then cancelling divergences would automatically appear
	in the coefficients of the resulting effective Lagrangian
	\cite{Burgess:1992gx,Arzt:1992wz}.
	For this reason, power law divergences such as those appearing
	in Eq.~\eref{eq:spectrum-divergences} are devoid of physical
	significance; only the coefficient of the logarithmic
	divergence can have meaning.
	If we begin with a Lorentz invariant effective low energy theory
	such as Eq.~\eref{eq:action} it is not possible to see these
	cancellations taking place.
	Thus, taken literally,
	our analysis up to this point is not compatible with a
	Lorentz violating cut-off; instead, sensible answers can be obtained
	only by using a Lorentz-invariant regulator such as dimensional
	reduction
	\cite{Weinberg:2005vy,Weinberg:2006ac,Chaicherdsakul:2006ui,
	Adshead:2008gk,Adshead:2009cb}.
	However, we can equally well make use of our knowledge that
	the power law terms in Eq.~\eref{eq:spectrum-divergences} must
	ultimately cancel, leaving behind an unfixed
	finite term or \emph{threshold correction}
	\cite{Georgi:1974yf,Weinberg:1978ym}.

	The threshold correction can be determined by matching to a more
	complete theory which resolves the details of ultraviolet physics,
	or by specifying a renormalization prescription which allows
	us to make contact with measurement.
	However, unlike simple
	theories such as quantum electrodynamics it is difficult to find
	an appropriate renormalization prescription in cosmology.
	This is because it is not possible to directly measure the
	expectation values we have computed: they are only important as an
	initial condition for the purpose of computing the structure in the
	late universe which \emph{is} visible to us.	
	A similar problem afflicts calculations in quantum chromodynamics,
	where interactions among hadrons are handled by first studying the
	predominantly \emph{electromagnetic} interactions among their
	constituent partons.%
		\footnote{Indeed, there is an interesting analogy between quantum
		chromodynamics and the calculation of cosmological expectation
		values using the $\delta N$ formalism. In QCD, one calculates
		correlation functions among partons at high energies, where the
		QCD coupling is small and perturbation theory applies.
		One then chooses a ``factorization scale'' which determines the
		energy below which partons are summed up into hadrons
		according to certain ``parton distribution functions.'' In doing
		so, it is possible to encounter large logarithmic singularities
		at soft or collinear
		momenta which enhance the phase space for interacting partons to
		dress themselves into jets, or which manifest as initial state
		radiation. One can find analogues for all these
		effects in cosmology: high energy corresponds to early times
		during inflation, where the slow-roll approximation applies and
		perturbation theory in slow-roll quantities makes sense;
		the time of horizon crossing plays the role of the factorization
		scale; the coefficients of the $\delta N$ expansion
		correspond to the parton distribution functions;
		and logarithms such as $N_\ast = \ln |k\eta_\ast|$
		play the role of the singularities at soft or collinear momenta.
		The details of this analogy have been explored in more detail
		in Ref.~\cite{Seery:2009hs}.}
	In the case of
	QCD it is also impossible to observe the parton correlation
	functions directly.
	
	Without extra information, little can be concluded about
	the magnitude of any possible threshold correction.
	If it is large it would correspond to a non-negligible mass for
	the inflaton, and therefore reproduces the well-known
	$\eta$-problem.
	Since it is necessary in any case
	to assume that some conspiracy of ultra-violet
	physics allows inflation to occur at all it is presumably
	a reasonable approximation to set the threshold correction to zero.
	We can then estimate a lower limit for the magnitude of the
	loop correction by supposing that the logarithm
	receives contributions up to the scale where
	renormalization begins to remove ultra-violet modes from the theory,
	leaving a contribution of the form
	$\ln (\Lambda'/\mu)$, where $\Lambda'$ is the scale
	of new physics.
	The nature of the cutoff is now immaterial. Taking
	$\mu \approx k \approx a_\ast H_\ast$ and $\Lambda'$ to be the
	proper scale of new physics, it follows that the loop corrected
	spectrum can be written
	\begin{equation}
		\fl
		P^{\alpha \beta}(k) = P_{0\ast}(k) \left\{
			\delta^{\alpha \beta} +
			\frac{4}{3} \Ps_{0\ast} \Omega_\ast^4
			\left( \frac{\lambda^\alpha \lambda^\beta}{\lambda^2} \right)_\ast
			(N_\ast + \ln 2 + \EulerMascheroni - 1)
			\ln \frac{\Lambda'}{H_\ast}
		\right\} ,
		\label{eq:one-loop-spectrum}
	\end{equation}
	where $P_0(k) = H^2 / 2k^2$ is the tree-level power spectrum,
	and $\Ps_0(k) = H^2 / 4\pi^2$ is its dimensionless equivalent.
	Note that since a flat metric is being assumed on field space, the
	placement of the $\alpha$ and $\beta$ field indices is immaterial.
	Eq.~\eref{eq:one-loop-spectrum} is the first principal result of this
	paper. It is interesting to observe that the logarithm
	$\ln (\Lambda'/H_\ast)$ cannot be too large;
	although the precise value of $H_\ast$ during inflation is model
	dependent, in chaotic models it is reasonable to assume that
	$H_\ast \approx 10^{-5}$ in fundamental units. It follows that
	$\ln (\Lambda'/H_\ast)$ cannot be more than of order $1$ -- $10$.
	
	The two-point function $\langle \delta\phi^\alpha(\vect{k}_1)
	\delta\phi^\beta(\vect{k}_2) \rangle$ is one contribution
	to the power spectrum $\zeta$, but in practice
	it would be accompanied by
	other contributions arising from non-linear terms in the gauge
	transformation between $\zeta$ and the $\delta\phi^\alpha$.
	These may themselves carry ultra-violet divergences which should be
	accounted for in an accurate calculation.
	However, there is no reason to believe that these contributions
	would be any larger than Eq.~\eref{eq:one-loop-spectrum},
	and we can therefore suppose that this expression suffices
	for the purpose of obtaining order-of-magnitude estimates.
	
	How large is the loop correction? Specializing for simplicity to
	the case of a single field, and adopting the parametrization
	$\lambda(\phi) = \exp(\phi/M_\phi)$, it follows that this loop correction
	does not overwhelm the tree-level provided $M_\phi^{-1} \lesssim 120$.
	This compares favourably with the value $M_\phi^{-1} \lesssim 20$,
	given in Eq.~\eref{eq:mass-scale}, which was suggested as appropriate
	for magnetogenesis (see \S\ref{sec:magnetogenesis};
	although $M_{\phi}^{-1}$ varies from model to model, it is unlikely
	to be as large as 120). Note that although
	these values may seem closer than desirable, the tuning here is in an
	exponential. A mass scale $M_\phi^{-1} = 120$ corresponds to a
	hierarchy $\lambda/\lambda_\ast = \e{1200}$ which is enormously
	larger than is required or desirable to produce a primordial
	magnetic field: the large energy density of an electromagnetic field
	amplified by such an enormous factor would swamp any other
	constituents of the universe and lead to an entirely unacceptable late
	time phenomenology.

	\section{Electric loop corrections to the scalar bispectrum}
	\label{sec:loop-bispectrum}
	
	Eq.~\eref{eq:one-loop-spectrum} and the associated bound on
	$\lambda$ or $M_\phi$ which guarantees that the loop correction does not
	overpower the tree-level are interesting in their own right.
	However,
	in this section we return to the question of non-gaussianity in the
	cosmic microwave background.
	One can obtain a
	marginally tighter bound on $M_\phi$ by demanding that the loop
	expansion is stable for this expectation value as well as for the
	spectrum, and we shall see that increasingly stringent bounds
	are possible for higher correlation functions.
	
\begin{figure}
		\begin{center}
			\hfill
			\begin{fmfgraph*}(60,60)
				\fmfpen{0.8thin}
				\fmftop{t}
				\fmfbottom{b1,b2}
				\fmf{plain,tension=3,label=$\scriptsize\vect{k}_1$,
				     label.side=left}{t,v1}
				\fmf{plain,tension=3,label=$\scriptsize\vect{k}_2$,
					 label.side=right}{b1,v2}
				\fmf{plain,tension=3,label=$\scriptsize\vect{k}_3$,
					 label.side=left}{b2,v3}
				\fmf{boson,side=left}{v1,v2}
				\fmf{boson,side=left}{v2,v3}
				\fmf{boson,side=left}{v3,v1}
				\fmfv{label=$\scriptsize +$,label.angle=145}{v1}
				\fmfv{label=$\scriptsize +$,label.angle=145}{v2}
				\fmfv{label=$\scriptsize +$,label.angle=35}{v3}
			\end{fmfgraph*}
			\hfill
			\begin{fmfgraph*}(60,60)
				\fmfpen{0.8thin}
				\fmftop{t}
				\fmfbottom{b1,b2}
				\fmf{plain,tension=3,label=$\scriptsize\vect{k}_1$,
				     label.side=left}{t,v1}
				\fmf{plain,tension=3,label=$\scriptsize\vect{k}_2$,
					 label.side=right}{b1,v2}
				\fmf{plain,tension=3,label=$\scriptsize\vect{k}_3$,
					 label.side=left}{b2,v3}
				\fmf{boson,side=left}{v1,v2}
				\fmf{boson,side=left}{v2,v3}
				\fmf{boson,side=left}{v3,v1}
				\fmfv{label=$\scriptsize +$,label.angle=145}{v1}
				\fmfv{label=$\scriptsize +$,label.angle=145}{v2}
				\fmfv{label=$\scriptsize -$,label.angle=35}{v3}
			\end{fmfgraph*}
			\hfill
			\begin{fmfgraph*}(60,60)
				\fmfpen{0.8thin}
				\fmftop{t}
				\fmfbottom{b1,b2}
				\fmf{plain,tension=3,label=$\scriptsize\vect{k}_1$,
				     label.side=left}{t,v1}
				\fmf{plain,tension=3,label=$\scriptsize\vect{k}_2$,
					 label.side=right}{b1,v2}
				\fmf{plain,tension=3,label=$\scriptsize\vect{k}_3$,
					 label.side=left}{b2,v3}
				\fmf{boson,side=left}{v1,v2}
				\fmf{boson,side=left}{v2,v3}
				\fmf{boson,side=left}{v3,v1}
				\fmfv{label=$\scriptsize +$,label.angle=145}{v1}
				\fmfv{label=$\scriptsize -$,label.angle=145}{v2}
				\fmfv{label=$\scriptsize +$,label.angle=35}{v3}
			\end{fmfgraph*}
			\hfill
			\begin{fmfgraph*}(60,60)
				\fmfpen{0.8thin}
				\fmftop{t}
				\fmfbottom{b1,b2}
				\fmf{plain,tension=3,label=$\scriptsize\vect{k}_1$,
				     label.side=left}{t,v1}
				\fmf{plain,tension=3,label=$\scriptsize\vect{k}_2$,
					 label.side=right}{b1,v2}
				\fmf{plain,tension=3,label=$\scriptsize\vect{k}_3$,
					 label.side=left}{b2,v3}
				\fmf{boson,side=left}{v1,v2}
				\fmf{boson,side=left}{v2,v3}
				\fmf{boson,side=left}{v3,v1}
				\fmfv{label=$\scriptsize -$,label.angle=145}{v1}
				\fmfv{label=$\scriptsize +$,label.angle=145}{v2}
				\fmfv{label=$\scriptsize +$,label.angle=35}{v3}
			\end{fmfgraph*}
			\hfill
			\mbox{}
		\end{center}
		\begin{center}
			\hfill
			\begin{fmfgraph*}(60,60)
				\fmfpen{0.8thin}
				\fmftop{t}
				\fmfbottom{b1,b2}
				\fmf{plain,tension=3,label=$\scriptsize\vect{k}_1$,
				     label.side=left}{t,v1}
				\fmf{plain,tension=3,label=$\scriptsize\vect{k}_2$,
					 label.side=right}{b1,v2}
				\fmf{plain,tension=3,label=$\scriptsize\vect{k}_3$,
					 label.side=left}{b2,v3}
				\fmf{boson,side=left}{v1,v2}
				\fmf{boson,side=left}{v2,v3}
				\fmf{boson,side=left}{v3,v1}
				\fmfv{label=$\scriptsize -$,label.angle=145}{v1}
				\fmfv{label=$\scriptsize -$,label.angle=145}{v2}
				\fmfv{label=$\scriptsize +$,label.angle=35}{v3}
			\end{fmfgraph*}
			\hfill
			\begin{fmfgraph*}(60,60)
				\fmfpen{0.8thin}
				\fmftop{t}
				\fmfbottom{b1,b2}
				\fmf{plain,tension=3,label=$\scriptsize\vect{k}_1$,
				     label.side=left}{t,v1}
				\fmf{plain,tension=3,label=$\scriptsize\vect{k}_2$,
					 label.side=right}{b1,v2}
				\fmf{plain,tension=3,label=$\scriptsize\vect{k}_3$,
					 label.side=left}{b2,v3}
				\fmf{boson,side=left}{v1,v2}
				\fmf{boson,side=left}{v2,v3}
				\fmf{boson,side=left}{v3,v1}
				\fmfv{label=$\scriptsize -$,label.angle=145}{v1}
				\fmfv{label=$\scriptsize +$,label.angle=145}{v2}
				\fmfv{label=$\scriptsize -$,label.angle=35}{v3}
			\end{fmfgraph*}
			\hfill
			\begin{fmfgraph*}(60,60)
				\fmfpen{0.8thin}
				\fmftop{t}
				\fmfbottom{b1,b2}
				\fmf{plain,tension=3,label=$\scriptsize\vect{k}_1$,
				     label.side=left}{t,v1}
				\fmf{plain,tension=3,label=$\scriptsize\vect{k}_2$,
					 label.side=right}{b1,v2}
				\fmf{plain,tension=3,label=$\scriptsize\vect{k}_3$,
					 label.side=left}{b2,v3}
				\fmf{boson,side=left}{v1,v2}
				\fmf{boson,side=left}{v2,v3}
				\fmf{boson,side=left}{v3,v1}
				\fmfv{label=$\scriptsize +$,label.angle=145}{v1}
				\fmfv{label=$\scriptsize -$,label.angle=145}{v2}
				\fmfv{label=$\scriptsize -$,label.angle=35}{v3}
			\end{fmfgraph*}
			\hfill
			\begin{fmfgraph*}(60,60)
				\fmfpen{0.8thin}
				\fmftop{t}
				\fmfbottom{b1,b2}
				\fmf{plain,tension=3,label=$\scriptsize\vect{k}_1$,
				     label.side=left}{t,v1}
				\fmf{plain,tension=3,label=$\scriptsize\vect{k}_2$,
					 label.side=right}{b1,v2}
				\fmf{plain,tension=3,label=$\scriptsize\vect{k}_3$,
					 label.side=left}{b2,v3}
				\fmf{boson,side=left}{v1,v2}
				\fmf{boson,side=left}{v2,v3}
				\fmf{boson,side=left}{v3,v1}
				\fmfv{label=$\scriptsize -$,label.angle=145}{v1}
				\fmfv{label=$\scriptsize -$,label.angle=145}{v2}
				\fmfv{label=$\scriptsize -$,label.angle=35}{v3}
			\end{fmfgraph*}
			\hfill
			\mbox{}
		\end{center}
		\caption{\label{fig:threepoint}Diagrams contributing to the
		three-point scalar expectation value. As before, straight lines
		(appearing on the external legs of the diagrams)
		indicate scalar quanta), whereas wavy lines indicate virtual
		gauge bosons circulating the the loop. These diagrams break
		into \emph{four} groups of complex conjugate pairs, with the
		$(+,+,+)$ and $(-,-,-)$ diagrams forming one pair and the
		permutations of the $(+,+,-)$ and $(-,-,+)$ diagrams
		forming the other three groups.}
	\end{figure}
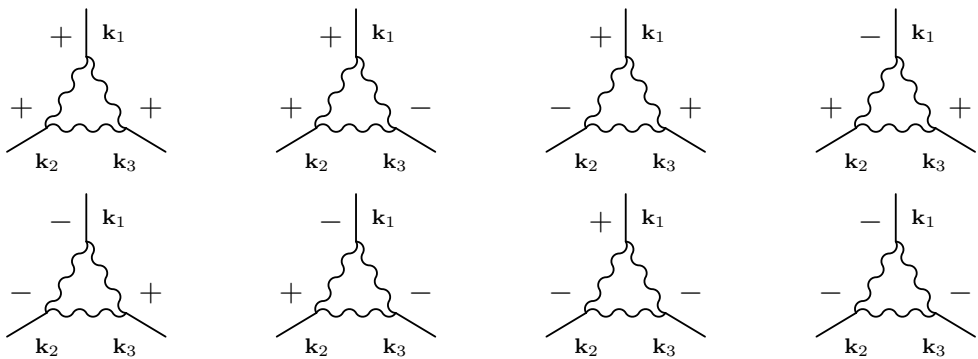

	The Schwinger formalism diagrams contributing to the three-point
	function are shown in Fig.~\ref{fig:threepoint}.
	These diagrams make contributions to the three-point function which
	can be put into a form similar to Eqs.~\eref{eq:twopoint-plusplus}
	and~\eref{eq:twopoint-plusminus} for the two-point function. For example,
	for the $(+,+,+)$ diagram we obtain
	\begin{eqnarray}
		\fl\nonumber
		\langle \delta\phi^\alpha(\vect{k}_1) \delta\phi^\beta(\vect{k}_2)
		\delta\phi^\gamma(\vect{k}_3) \rangle_\ast \supseteq
		\frac{H_\ast^6}{64\prod_i k_i^3}
		\left( \frac{\lambda_\alpha \lambda_\beta \lambda_\gamma}
			{\lambda^3}
		\right)_\ast
		\Omega_\ast^6
		\prod_i (1 + \im k_i \eta_\ast) \e{- \im k_i \eta_\ast}
		\\ \nonumber \mbox{}
		\times
		\int \frac{\d^3 q \, \d^3 r \, \d^3 s}{q^3 r^3 s^3}
		P_3(\vect{q},\vect{r},\vect{s})
		\delta(-\vect{k}_1 - \vect{r} + \vect{q})
		\delta(-\vect{k}_2 - \vect{s} + \vect{r})
		\delta(-\vect{k}_3 - \vect{q} + \vect{s}) \\ \mbox{} \times
		Q_+(k_1,k_2,k_3,q,r,s) ,
		\label{eq:threepoint-plusplus}
	\end{eqnarray}
	where now $i \in \{ 1, 2, 3 \}$ and
	$Q_+$ is a function of the scalar momenta which is defined by
	\begin{eqnarray}
		\fl\nonumber
		Q_+ \equiv
		\threeEi{k_1}{k_2}{k_3}{k_1+q+r}{k_2-r+s}{k_3-q-s} +
		\threeEi{k_1}{k_3}{k_2}{k_1+q+e}{k_3-q+s}{k_2-r-s} \\ \nonumber
		\mbox{} +
		\threeEi{k_2}{k_1}{k_3}{k_2+r+s}{k_1+q-r}{k_3-q-s} +
		\threeEi{k_3}{k_1}{k_2}{k_3+q+s}{k_1-q+r}{k_2-r-s} \\
		\mbox{} +
		\threeEi{k_2}{k_3}{k_1}{k_2+r+s}{k_3+q-s}{k_1-q-r} +
		\threeEi{k_3}{k_2}{k_1}{k_3+q+s}{k_2+r-s}{k_1-q-r} .
		\label{}
	\end{eqnarray}
	The momentum polynomial $P_3(\vect{q},\vect{r},\vect{s})$
	is the analogue of Eq.~\eref{eq:spectrum-polynomial} for the bispectrum,
	and depends symmetrically on each of the vector momenta
	$\vect{q}$, $\vect{r}$ and $\vect{s}$. Specifically, it obeys
	\begin{equation}
		P_3(\vect{q},\vect{r},\vect{s}) \equiv
			q^2(\vect{r}\cdot\vect{s})^2 +
			r^2(\vect{q}\cdot\vect{s})^2 +
			s^2(\vect{q}\cdot\vect{r})^2 -
			(\vect{q}\cdot\vect{r})
			(\vect{q}\cdot\vect{s})
			(\vect{r}\cdot\vect{s}) ,
	\end{equation}
	whereas
	the time integrals can now be cast in the form of a
	six-parameter exponential integral,
	\begin{equation}
		\fl
		\threeEi{\alpha_1}{\alpha_2}{\alpha_3}
			{\beta_1}{\beta_2}{\beta_3} \equiv
			- \im \int_{-\infty}^{\eta_\ast} \frac{\d \zeta}{\zeta^2}
			\int_{-\infty}^{\zeta} \frac{\d \tau}{\tau^2}
			\int_{-\infty}^{\tau} \frac{\d \eta}{\eta^2}
			(1 - \im \alpha_1 \eta)(1 - \im \alpha_2 \tau)
			(1 - \im \alpha_3 \zeta)
			\e{\im \beta_1 \eta} \e{\im \beta_2 \tau} \e{\im \beta_3 \zeta} ,
		\label{eq:time-three}
	\end{equation}
	with the necessary deformations of the contour of integrations---or,
	more accurately, the inclusion of suppression factors in the
	exponentials---to cause convergence at large $|\eta|$, $|\tau|$ or
	$|\zeta|$.
	We expect the same general considerations to govern the result of
	the integrals in Eq.~\eref{eq:threepoint-plusplus} which applied
	for the two-point function: after integrating by parts,
	Eq.~\eref{eq:time-three} can be expressed as an asymptotic series
	in inverse powers of $\eta_\ast$, together with nested
	integrals of the form $\int \d \zeta \, \zeta^{-1} \,
	\e{\im \mu \zeta}$ for some $\mu$.
	
	We must also take account of the $(+,+,)$- and $(-,+,+)$-type diagrams.
	Since these are related by complex conjugation it suffices to
	consider only diagrams of one particular signature, for which we
	will choose the $(+,+,-)$-type, and then add in their complex
	conjugates. The $-$ vertex can be attached to
	any of the external legs, corresponding to the any one
	of the momenta $\vect{k}_i$.
	The contribution from any such diagram, for example where the
	$-$ vertex is associated with $\vect{k}_3$, can be written
	in a form similar to Eq.~\eref{eq:threepoint-plusplus},
	taking account of the fact that, irrespective of its numerical
	value, the time at the $-$ vertex is taken to be later than the
	time of evaluation $\eta_\ast$, and with
	$Q_+$ replaced by a different function $Q_-$ which accounts for
	the necessary sign interchanges. We therefore arrive at
	\begin{eqnarray}
		\fl\nonumber
		\langle \delta\phi^\alpha(\vect{k}_1)
		\delta\phi^\beta(\vect{k}_2)
		\delta\phi^\gamma(\vect{k}_3) \rangle \supseteq
		\frac{H_\ast^6}{64 \prod_i k_i^3}
		\left(
			\frac{\lambda_\alpha \lambda_\beta \lambda_\gamma}{\lambda^3}
		\right)_\ast
		\Omega_\ast^6
		\\ \nonumber \mbox{} \times
		(1 + \im k_1 \eta_\ast) (1 + \im k_2 \eta_\ast)
		(1 - \im k_3 \eta_\ast) \e{-\im \eta_\ast (k_1 + k_2 - k_3)}
		\\ \nonumber \mbox{} \times
		\int \frac{\d^3 q \, \d^3 r \, \d^3 s}{q^3 r^3 s^3}
		P_3(\vect{p},\vect{q},\vect{r})
		\delta(-\vect{k}_1-\vect{r}+\vect{q})
		\delta(-\vect{k}_2-\vect{s}+\vect{r})
		\delta(-\vect{k}_3-\vect{q}+\vect{s})
		\\ \mbox{} \times
		Q_-(k_1,k_2,k_3,q,r,s) ,
	\end{eqnarray}
	together with the equivalent expressions obtained by exchanging
	$\vect{k}_1$ with $\vect{k}_2$ and $\vect{k}_3$.
	In this expression, $Q_-$ is defined by
	\begin{eqnarray}
		\fl\nonumber
		Q_- \equiv
			\threeEi{k_1}{k_2}{-k_3}{k_1+q+r}{k_2-r+s}{-k_3-q-s} +
			\threeEi{k_2}{k_1}{-k_3}{k_2+r+s}{k_1+q-r}{-k_3-q-s}
			\\ \nonumber \mbox{} +
			\threeEi{k_1}{-k_3}{k_2}{k_1+q+r}{-k_3-q-s}{k_2-r+s} +
			\threeEi{k_2}{-k_3}{k_1}{k_2+r+s}{-k_3-q-s}{k_2+q-r}
			\\ \mbox{} +
			\threeEi{-k_3}{k_1}{k_2}{-k_3-q-s}{k_1+q+r}{k_2-r+s} +
			\threeEi{-k_3}{k_2}{k_1}{-k_3-q-s}{k_2+r+s}{k_1+q-r} .
	\end{eqnarray}
	
	To see how this works in detail, it follows after integrating by parts
	that we can express Eq.~\eref{eq:time-three} in the form
	\begin{eqnarray}
		\fl\nonumber
		\threeEi{\alpha_1}{\alpha_2}{\alpha_3}
			{\beta_1}{\beta_2}{\beta_3} =
		\left( \frac{\im}{6 \eta_\ast^3} + \frac{\xi_1}{\eta_\ast^2}
			+ \frac{\im \xi_2}{\eta_\ast}
		\right) \e{\im(\beta_1 + \beta_2 + \beta_3) \eta_\ast} +
		(\beta_1 + \beta_2 + \beta_3) \xi_2 \int_{-\infty}^{\eta_\ast}
			\frac{\d \zeta}{\zeta} \e{\im (\beta_1 + \beta_2 + \beta_3)
			\zeta} \\ \nonumber \mbox{}
		- \frac{\im \xi_3}{\eta_\ast} \e{\im \beta_3 \eta_\ast}
			\int_{-\infty}^{\eta_\ast} \frac{\d \zeta}{\zeta}
			\e{\im(\beta_1 + \beta_2) \zeta}
		+ \left( \frac{\xi_4}{2\eta_\ast^2} -
			\frac{\im \xi_5}{\eta_\ast} \right) \e{\im (\beta_2 + \beta_3)
			\eta_\ast} \int_{-\infty}^{\eta_\ast} \frac{\d \zeta}
			{\zeta} \e{\im \beta_1 \zeta} \\ \nonumber \mbox{}
		- \xi_6 \int_{-\infty}^{\eta_\ast} \frac{\d \zeta}{\zeta}
			\e{\im \beta_3 \zeta} \int_{-\infty}^{\zeta}
			\frac{\d \tau}{\tau} \e{\im (\beta_1 + \beta_2) \tau}
		- \xi_7 \int_{-\infty}^{\eta_\ast} \frac{\d \zeta}{\zeta}
			\e{\im (\beta_2 + \beta_3) \zeta} \int_{-\infty}^{\zeta}
			\frac{\d \tau}{\tau} \e{\im \beta_1 \tau} \\ \nonumber \mbox{}
		- \frac{\im \xi_8}{\eta_\ast} \e{\im \beta_3 \eta_\ast}
			\int_{-\infty}^{\eta_\ast} \frac{\d \zeta}{\zeta}
			\e{\im \beta_2 \zeta} \int_{-\infty}^{\zeta}
			\frac{\d \tau}{\tau} \e{\im \beta_1 \tau} \\ \mbox{}
		- \xi_9 \int_{-\infty}^{\eta_\ast} \frac{\d \zeta}{\zeta}
			\e{\im \beta_3 \zeta} \int_{-\infty}^{\zeta}
			\frac{\d \tau}{\tau} \e{\im \beta_2 \tau}
			\int_{-\infty}^{\tau} \frac{\d \eta}{\eta}
			\e{\im \beta_1 \eta} ,
		\label{eq:time-three-asymptotic}
	\end{eqnarray}
	where the coefficients $\xi_1$ to $\xi_9$ are defined by
	\begin{eqnarray}
		\xi_1 \equiv \frac{1}{2} \Big\{
			9 (\alpha_2 - \alpha_1) + 3 \alpha_3 +
			2 (\beta_1 - 2\beta_2) - \beta_3
		\Big\} \\ \nonumber
		\xi_2 \equiv \frac{1}{2} \Big\{
			9 \alpha_3 \beta_2 + 2 \beta_2^2 - 3 \alpha_3 \beta_1 -
			4 \beta_1^2 - 2 \beta_1 \beta_2 - \beta_3^2 +
			3 \alpha_1 (8 \alpha_3 + \beta_1 - \beta_2 - 5 \beta_3)
			\\ \quad \quad \quad \quad \mbox{} +
			\beta_3 (3 \alpha_3 + \beta_1 - 5 \beta_2 ) +
			3 \alpha_2 (\beta_2 - 8 \alpha_3 - \beta_1 + 5 \beta_3)
		\Big\} \\
		\xi_3 = (\beta_1 + \beta_2) \left(
			\alpha_2 - \alpha_2 - \frac{\beta_2 - \beta_1}{2}
		\right) \\
		\xi_4 = \beta_1 - \alpha_1 \\
		\xi_5 = \frac{1}{2} (\beta_1 - \alpha_1)
			(2 \alpha_2 + 2\alpha_3 - 3 \beta_2 - \beta_3) \\
		\xi_6 = (\beta_3 - \alpha_3)(\beta_1 + \beta_2)
			\left( \alpha_2 - \alpha_1 - \frac{\beta_2 - \beta_1}{2}
			\right) \\
		\xi_7 = (\beta_2 + \beta_3) \Bigg\{
			\alpha_3 + (\beta_1 - \alpha_1)(\beta_2 - \alpha_1)
			- \frac{\beta_2}{2} - \frac{\beta_3}{2}
			\Bigg\} \\
		\xi_8 = (\beta_1 - \alpha_1)(\beta_2 - \alpha_2) \\
		\xi_9 = (\beta_1 - \alpha_1)(\beta_2 - \alpha_2)
			(\beta_3 - \alpha_3)
	\end{eqnarray}
	This function contains primitive fast divergences, but as in the
	case of the spectrum these will conspire to cancel among themselves
	in the final answer, leaving a result which contains only slow
	divergences which scale as powers of $N_\ast$.
	As before, the combination $\beta_1 + \beta_2 + \beta_3$ is
	always independent of the loop momentum, whereas most of the integrals
	involve decaying exponentials of $|\beta_1|$, $|\beta_2|$ or $|\beta_3|$
	and will play no role when the loop integral is dominated by its
	ultra-violet region. In $Q_+$, there is no combination other than
	$\beta_1 + \beta_2 + \beta_3$ which can contribute, whereas
	in $Q_-$---for four of the terms---the combination $\beta_1 + \beta_2$
	(and therefore $\beta_3$ on its own) \emph{also}
	remains finite in the extreme ultra-violet limit.
	It follows that the terms involving $\xi_3$ and $\xi_6$ can also
	contribute, although it will turn out that in
	Eq.~\eref{eq:time-three-asymptotic} all these pieces cancel
	out of the final answer.

	In order to simplify the calculation from this point onwards,
	it is convenient to specialize
	to the equilateral limit in which all the $k_i$ are equal,
	giving $k_1 = k_2 = k_3$. In practice this does not entail much loss
	of generality, since the approximations we are obliged to make
	in evaluating the time integrals in
	Eqs.~\eref{eq:time-two} and~\eref{eq:time-three} mean that we cannot
	take the $k_i$ to exit the horizon with a separation of more than
	a few e-folds. (Exactly similar remarks apply to the standard
	calculation of the tree-level three-point function.)
	One may now proceed by analogy with the scalar two-point function,
	as described in \S\ref{sec:loop-spectrum}. The three-point function
	is written in the form
	\begin{equation}
		\fl
		\langle \delta\phi^\alpha(\vect{k}_1)
		\delta\phi^\beta(\vect{k}_2)
		\delta\phi^\gamma(\vect{k}_3)
		\rangle_{\ast,k_i = k} =
		(2\pi)^3 \delta(\sum_i \vect{k}_i)
		\frac{H_\ast^6}{64 \prod_i k_i^3}
		\left(
			\frac{\lambda^\alpha \lambda^\beta \lambda^\gamma}
			{\lambda^3}
		\right)_\ast
		\Omega_\ast^6
		I_3 ,
	\end{equation}
	and the time- and momentum-dependence is absorbed into $I_3$.
	This integral is somewhat more complicated than its counterpart
	for the two-point function. It is defined by
	\begin{eqnarray}
		\fl\nonumber
		I_3 \equiv 4k \int \frac{\d^3 q}{q^3 r^3 s^3}
		\; P_3(\vect{q},\vect{r},\vect{s})
		\Big\{
			4k(q+r+s) + (q^2 + r^2 + s^2) \ln 27
			\\ \mbox{} \qquad \qquad \qquad + (qr + qs + rs)
			(4 N_\ast + 4 \EulerMascheroni + 9 \ln 3 - 4) - 2k^2
		\Big\} ,
	\end{eqnarray}
	with the specific assignments $\vect{r} \equiv \vect{q} - \vect{k}_1$
	and $\vect{s} \equiv \vect{q} + \vect{k}_3$. In addition, $N_\ast$
	satisfies the usual definition $N_\ast \equiv \ln |k \eta_\ast|$, which
	is unambiguous here because of our assumption that the momentum triangle
	formed by the $\vect{k}_i$ is equilateral.
	
	To proceed, we pick $\vect{k}_1$ to point along the
	$\hat{\vect{z}}$ axis of a spherical
	polar coordinate system. (We could also pick $\vect{k}_3$, but the
	result is independent of our choice.) In this system of coordinates
	the only non-trivial inner product we require is that of
	$\vect{q}$ with $\vect{k}_3$, which takes the form
	\begin{equation}
		\vect{q} \cdot \vect{k}_3 =
		kq \Big\{
			\cos \theta \cos \theta_{13} + \sin \theta \sin \theta_{13}
			\cos ( \phi - \phi_{13} )
		\Big\} ,
	\end{equation}
	where $\{ \theta, \phi \}$ are the zenith and azimuth of
	$\vect{q}$ relative to $\vect{k}_1$, and
	$\{ \theta_{13}, \phi_{13} \}$ are the corresponding zenith and azimuth 
	of $\vect{k}_{13}$. Note that $\phi_{13}$ is devoid of
	significance, being only a coordinate choice, and cannot appear in
	any physical quantity, whereas $\theta$ and $\phi$ are simply
	variables of integration. The integral is quadratically divergent.
	With a cut-off $\Lambda$, we find
	\begin{equation}
		\fl
		I_3 = - \frac{24 k}{\pi^2} \Lambda^2
			\left( \tilde{N}_\ast - \frac{1}{4} \ln 3 \right)
			- \frac{48 k^2}{\pi^2} \Lambda
			+ \frac{8 k^3}{\pi^2} \ln \frac{\Lambda}{\mu}
			\tilde{N}_\ast
			\left(
				4 + 2 \cos \theta_{13} +
				\frac{1}{\tilde{N}_\ast}
			\right) ,
		\label{eq:i-three}
	\end{equation}
	where $\tilde{N}_\ast$ is a slightly modified count of the number of
	e-folds since horizon exit, defined by
	\begin{equation}
		\tilde{N}_\ast \equiv N_\ast + \EulerMascheroni + \frac{11}{4}
		\ln 3 - 1 \approx N_\ast + 2.60 .
	\end{equation}
	One might worry that the appearance of $\theta_{13}$ in
	Eq.~\eref{eq:i-three} represents of failure of symmetry between
	$\vect{k}_1$, $\vect{k}_2$ and $\vect{k}_3$. However, it should be
	remembered that once we specify any two of the momenta the third
	is determined by the triangle law $\vect{k}_1 + \vect{k}_2 +
	\vect{k}_3 = 0$. Only the relative orientations of these vectors
	can enter physical quantities, and the absolute orientation of
	\emph{one} vector and the azimuth of a second are simply coordinate
	choices. For an equilateral configuration, this leaves only
	\emph{one} cosine which can play a role in physical quantities.
	
	The tree-level,
	single-field equilateral bispectrum is known to take the form
	\cite{Seery:2005gb,Seery:2008qj}
	\begin{equation}
		B_{0 \ast} = - \frac{11}{2} \sqrt{2 \epsilon_\ast}
		\frac{H_\ast^4}{2k^6} .
	\end{equation}
	We drop power law divergences and neglect a possible threshold
	correction.
	This implies that, if we specialize to single-field inflation
	and return to the parametrization $\lambda(\phi) = \exp ( \phi /
	M_\phi)$, the one-loop corrected bispectrum can be written
	as
	\begin{equation}
		B_\ast \approx B_{0 \ast} \left( 1 - \frac{4}{11 \sqrt{2}}
		\frac{\epsilon_\ast^{7/2}}{M_\phi^9} \Ps_{\R} \tilde{N}_\ast
		\ln \frac{\Lambda'}{H_\ast} \right) ,
	\end{equation}
	where $\Lambda'$ is again the proper scale of new physics,
	which can have a hierarchy with respect to the Planck
	scale of no more than around $\ln (\Lambda'/H_\ast) \sim 10$.
	The power spectrum of the curvature perturbation, $\Ps_{\R}$,
	is computed using the tree-level scalar spectrum, which we assume is
	not destabilized by the loop correction from electric quanta.
	It follows that if, likewise,
	we do not wish the loop correction to be larger than
	the tree-level, we must have $M_\phi^{-1} \lesssim 50$.
	
	This bound is
	somewhat stronger than is required to guarantee stability of the
	spectrum. Indeed, it is clear that successively higher-order expectation
	values will contain increasingly large negative powers of $M_\phi$,
	whereas the leading correction to an $n$-point scalar expectation
	value comes from diagrams such as those in Fig.~\ref{fig:threepoint}
	where the electromagnetic quanta circulate in a ring between the
	external scalar legs. Such a diagram is always suppressed with
	respect to the tree-level by only one power of the loop-counting
	parameter $(H/\Planck)^2$, and therefore for very large $n$ we will
	require $M_\phi$ to be closer and closer to unity.
	Does this imply that only $M_\phi = 1$ makes sense for an effective
	field theory? In fact, this conclusion would be too strong.
	For $M_\phi$ too far below the Planck scale the present analysis
	does not show that the two- and three-point scalar expectation values
	are inconsistent with observation, but rather only that perturbation
	theory is insufficient to compute them.
	
	It is known that the tree-level power spectrum is an excellent match
	for observation, and that the temperature anisotropy is gaussian to
	high accuracy. It follows that there is a reasonably secure foundation
	to demand that the two-point function can be computed perturbatively.
	The situation for the bispectrum is less clear. It is known that
	the non-linearity of the temperature anisotropy is not too large,
	although it may perhaps be of order $\fnl \sim 60$ in the
	squeezed limit. This limit is the opposite of the configuration
	studied here, where one of the momenta is going to zero while the other
	two become approximately equal and opposite, but even in the
	equilateral case where estimation is more difficult there is a
	relatively stringent
	limit, $|\fnl| \lesssim 300$. There is not yet any hint from the data
	that the three-point function requires anything beyond conventional
	perturbation theory.
	Even less is known concerning the
	properties of the $n$-point expectation values for $n$ larger than
	three. For this reason, it seems most conservative to adopt the
	approximate bound $M_{\phi}^{-1} \lesssim 120$ which corresponds to
	the applicability of perturbation theory for the power spectrum.
	
	\section{Discussion}
	\label{sec:conclude}
	
	In this paper I have computed the corrections to the spectrum and
	bispectrum of a set of light scalar fields which arise from coupling
	to an electromagnetic field during inflation.
	The calculation of inflationary observables is conventionally split
	into a quantum-mechanical initial condition, which specifies the
	momentum-dependence of expectation values at the time of horizon crossing,
	and a subsequent classical evolution.
	The interaction studied in this paper constitutes a correction to
	the initial condition, and can be thought of as arising from
	interactions with virtual quanta of the electromagnetic field.
	In a scenario where the gauge coupling is rolling rapidly during
	inflation, these virtual electromagnetic quanta experience strong
	amplification as they are drawn across the horizon and ``freeze in''
	to the correlations among the scalar modes at that time.
	
	To which scenarios would this correction apply? The most direct
	application is to theories of so-called
	magnetogenesis, where one aims to
	produce a late-time magnetic field by breaking the conformal
	invariance of the $U(1)$ gauge field.
	In this scenario the scalar fields which determine the
	expectation value $\lambda(\phi)$ must be rolling on cosmological
	timescales, meaning that these scalars will typically also contribute
	to the curvature perturbation, $\zeta$. It is $\zeta$ which is
	visible as the adiabatic temperature perturbation in the
	Cosmic Microwave Background. Therefore, in order that the conventional
	perturbative predictions for the power spectrum and
	bispectrum are not destabilized we must demand that the correction
	from electromagnetic interactions does not dominate the tree-level.
	As shown in \S\ref{sec:loop-spectrum}, this requires
	that $M_{\phi}^{-1} \lesssim 120$ in order that the spectrum is
	not destabilized, and the stronger condition
	(\S\ref{sec:loop-bispectrum})
	$M_{\phi}^{-1} \lesssim 50$ for the bispectrum. Both of these
	conditions are approximately satisfied for conventional pictures
	of magnetogenesis, which,
	as shown in \S\ref{sec:magnetogenesis},
	typically requires $M_{\phi}^{-1} \approx 20$.
	According to the discussion
	in \S\ref{sec:loop-bispectrum} it is probably most
	conservative to adopt the limit $M_{\phi}^{-1} \lesssim 120$,
	because the evidence that current observation matches perturbative 
	calculations of the bispectrum is rather less strong than for the
	spectrum.
	
	There is another mechanism by which synthesis of an electric or
	magnetic field could destabilize conventional perturbation theory.
	If the energy density in electromagnetic radiation begins to compete
	with the potential energy associated with the scalar fields, then
	this would deform the comoving hypersurface on which inflation ends.
	This deformation would imply that $\zeta$ was not conserved,
	but rather would acquire a contribution from this final hypersurface
	\cite{Lyth:2005qk}.	
	On the other hand,
	if the energy density in the electromagnetic field remains small,
	then we can be confident that there is no back-reaction on the
	metric. In this case,
	it follows that the hypersurface on which inflation ends
	suffers a negligible deformation
	and $\zeta$ is given by the usual procedure, because the electromagnetic
	energy density never contributes to the expansion rate or the
	integrated number of e-foldings, $N$.
	
	When does energy in the electromagnetic field remain small? This
	question was addressed by Bamba \cite{Bamba:2006ga}
	and Martin \& Yokoyama \cite{Martin:2007ue}, who gave a detailed
	discussion of the back-reaction problem. They concluded that
	when a magnetic field is synthesized there is never a problem with
	back reaction, whereas if the outcome is an electric field
	then the scenario is only free of instabilities associated with
	back reaction if the scale of inflation is taken to be very low.
	One way to understand
	this phenomenon is to note that the magnetic energy density
	$\rho_B \sim (\partial \omega)^2$ is proportional to
	spatial gradients,
	whereas the electric energy density $\rho_E \sim \dot{\omega}^2$
	involves a time derivative. We must therefore include
	$\rho_E$ at leading order in a gradient expansion of the scalar
	Klein--Gordon equation. If $\rho_E$ becomes too large
	it begins to act as a source and causes
	an unwanted growing instability in the scalar fluctuations.
	In practice this ``instability'' may entail nothing more
	dramatic than
	a macroscopic flow of charge which shorts out the electric field
	\cite{Bamba:2006ga}, but our ability to calculate is impaired
	and it is difficult to draw conclusions with any confidence.
	Therefore, unless the scale of inflation is taken to be very low,
	it seems that one should restrict attention to scenarios
	where $\lambda(\phi)$ grows during inflation.
		
	Couplings between scalars and $U(1)$ gauge fields are invoked in
	certain theories of dark energy, such as the so-called ``chameleon''
	\cite{Khoury:2003aq} model. Such couplings were
	proposed in Ref.~\cite{Brax:2007hi},
	following earlier work
	\cite{Mota:2003tm,Clifton:2004st},
	and studied using astrophysical
	means in Refs.~\cite{Burrage:2007ew,Burrage:2008ii,Burrage:2009mj}.
	(For laboratory limits, see Refs.~\cite{Chou:2008gr,Afanasev:2008jt}.)
	However, the chameleon is essentially inert during inflation: it
	rolls rapidly
	to its minimum, and is fixed there for the duration of inflation
	\cite{Brax:2004qh}.
	Since it is not in motion while observable scales are leaving the
	horizon,
	the parameter $\Omega$ which controls the
	magnitude of the correction tends to zero.
	Moreover,
	because the energy density carried by the chameleon is
	cosmologically negligible,
	its expectation values do not contribute to those of the curvature
	perturbation. Therefore,
	although perturbation theory may fail for expectation values of the
	chameleon field if the associated mass scale is too small, this has
	no observational consequences. We conclude that no limit on the
	chameleon coupling $\beta$ analogous to those obtained in
	Refs.~\cite{Brax:2007hi,Burrage:2007ew,Burrage:2008ii} can be
	extracted from the present analysis.
	
	Although fast instabilities were encountered at intermediate points
	in the calculation, the final expectation values were found to be
	free of fast divergences. Their time dependence was instead described
	by powers of $N_\ast$, which measures by how many e-folds the
	fluctuation in question has passed outside the horizon.
	One can understand this
	as a consequence of a theorem due to Weinberg, which guarantees
	that interactions involving gauge fields are well behaved in this
	sense \cite{Weinberg:2006ac}. In the present case, although the
	electromagnetic part of the interaction would be ``dangerous''
	(in Weinberg's sense) as part of a scalar interaction because it
	involves time derivatives, it does not
	produce fast divergences. Indeed, the only effect of differentiation
	with respect to time when applied to a mode of the gauge field
	is to introduce factors of the slowly varying parameter $\Omega$.
	Such factors merely set the scale of the interaction, rather than
	changing the character of the time dependence, as would be the case
	for a scalar mode.
	
	\ack
	
	I would like to thank
	Clare Burrage, Anne Davis, Jim Lidsey, David Lyth, Karim Malik,
	Amanda Weltman and Daniel Wesley, who helped me with a large number of
	details concerning cosmological magnetic fields.
	Peter Adshead, Emanuela Dimastrogiovanni, Richard Easther,
	Eugene Lim, Meindert van der Meulen, Sarah Shandera, Jan Smit,
	Andrew Tolley and Daniel Wesley
	made many important suggestions concerning the treatment of loop diagrams
	in de Sitter space.
	
	I would like to thank the Astronomy Unit at Queen Mary, University of
	London, for their hospitality.	
	
	\appendix
		
	\section{Scalar loop corrections from the $V'''$ vertex}
	\label{appendix:loop}
	
\begin{figure}
		\begin{center}
			\begin{fmfgraph*}(80,60)
				\fmfleft{l}
				\fmfright{r}
				\fmf{plain,label=$\vect{k}_1$,tension=3,
				     label.side=left}{l,v1}
				\fmf{plain,label=$\vect{k}_2$,tension=3,
				     label.side=right}{r,v2}
				\fmf{fermion,left}{v1,v2}
				\fmf{fermion,left}{v2,v1}
				\fmfv{label=$V'''$,label.angle=-120}{v1}
				\fmfv{label=$V'''$,label.angle=-60}{v2}
			\end{fmfgraph*}
		\end{center}
		\caption{\label{fig:v-vertex}All-scalar loop correction from
		the $V'''$ vertex, common to all scalar field models of inflation.}
	\end{figure}
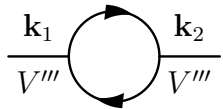

	In this Appendix, I compute the scalar loop correction coming from
	a simple $V'''$ vertex. This loop correction is common to any
	inflationary model, and arises simply from the self-interactions
	implied by the scalar potential, unless $V'''$ is somehow
	tuned to be zero.
	This loop correction corresponds to the diagram in
	Fig.~\ref{fig:v-vertex}, with two external scalar legs which are
	connected by a loop of circulating virtual scalar quanta.
	Just like the processes considered in the main text,
	this diagram comes in four different
	types labelled by the $+$ or $-$ flavours at each vertex,
	giving complex conjugate pairs $(+,+)$, $(-,-)$ and
	$(+,-)$, $(-,+)$.
	
	The loop process described by Fig.~\ref{fig:v-vertex} has already
	appeared in the literature, and has been the subject of detailed
	study by several previous authors. van der Meulen \& Smit
	\cite{vanderMeulen:2007ah} studied this loop for both large and small
	internal momenta, finding corrections to the classical time
	dependence. These corrections are one possible source of the
	``quantum logarithms'' described in the Introduction
	(\S\ref{sec:introduction}).
	Bartolo {\etal} \cite{Bartolo:2007ti}
	considered the same loop process, but focused
	on the infra-red region and found results in this regime which were
	in agreement with those
	of van der Meulen \& Smit. In this Appendix the same loop is studied
	using the methodology applied in the main text, principally
	for the purpose of aiding comparison of the methods of the present paper
	with those of either van der Meulen \& Smit or Bartolo {\etal}
	
	We work with a single-field model of inflation. The generalization
	to multiple fields is easy to obtain by identical methods,
	and keeping track of the necessary indices in the space of scalar fields
	only introduces needless notational clutter.
	It follows that the vertex is simple, and takes the form
	\begin{equation}
		S_3 \supseteq -
		\int \frac{\d^3 k_1 \, \d^3 k_2 \, \d^3 k_3}{(2\pi)^9}
		\; (2\pi)^3 \delta( \sum_i \vect{k}_i )
		\frac{V'''}{3} \delta\phi(\vect{k}_1)
		\delta\phi(\vect{k}_2)
		\delta\phi(\vect{k}_3) .
	\end{equation}
	As with the two-point loop correction studied in
	\S\ref{sec:loop-spectrum}, it is sufficient to compute the contribution
	from the $(+,+)$- and $(+,-)$-type diagrams and then add in their
	complex conjugates. From the $(+,+)$ diagram one obtains
	\begin{eqnarray}
		\fl\nonumber
		\langle \delta\phi(\vect{k}_1) \delta\phi(\vect{k}_2) \rangle_\ast
		\supseteq - \frac{1}{8 k_1^3 k_2^3} (V_\ast''')^2
		\int \frac{\d^3 q \, \d^3 r}{q^3 r^3}
		\delta(-\vect{k}_1 + \vect{q} - \vect{r})
		\delta(-\vect{k}_2 - \vect{q} + \vect{r})
		\\ \nonumber \mbox{} \times
		(1 + \im k_1 \eta_\ast) (1 + \im k_2 \eta_\ast)
		\e{-\im \eta_\ast (k_1 + k_2)}
		\\ \mbox{} \times
		\left\{ \threeEiVariant{k_1-q-r}{k_1 q + k_1 r - q r}{k_1 q r}
							   {k_2+q+r}{-k_2 q - k_2 r - q r}{k_2 q r} +
				\threeEiVariant{k_2-q-r}{k_2 q + k_2 r - q r}{k_2 q r}
							   {k_1+q+r}{-k_1 q - k_1 r - q r}{k_1 q r}
		\right\} ,
	\end{eqnarray}
	whereas from the $(+,-)$ diagram one obtains
	\begin{eqnarray}
		\fl\nonumber
		\langle \delta\phi(\vect{k}_1) \delta\phi(\vect{k}_2) \rangle_\ast
		\supseteq \frac{1}{8 k_1^3 k_2^3} (V_\ast''')^2
		\int \frac{\d^3 q \, \d^3 r}{q^3 r^3}
		\delta(-\vect{k}_1 + \vect{q} - \vect{r})
		\delta(-\vect{k}_2 - \vect{q} + \vect{r})
		\\ \nonumber \mbox{} \times
		(1 + \im k_1 \eta_\ast) (1 - \im k_2 \eta_\ast)
		\e{- \im \eta_\ast (k_1 - k_2)}
		\\ \nonumber\mbox{} \times
		\left\{ \threeEiVariant{k_1+q+r}{-k_1 q - k_1 r - q r}{k_1 q r}
							   {-k_2-q-r}{-k_2 q - k_2 r - q r}{-k_2 q r} +
				\threeEiVariant{-k_2-q-r}{-k_2 q - k_2 r - q r}{-k_2 q r}
							   {k_1+q+r}{-k_1 q - k_1 r - q r}{k_1 q r}
		\right\} .
		\\
	\end{eqnarray}
	As in \S\ref{sec:loop-spectrum}, these contributions are symmetric
	between $\vect{q}$ and $\vect{r}$, and are both written
	in terms of a time integral which takes the form
	\begin{eqnarray}
		\fl\nonumber
		\threeEiVariant{\alpha_1}{\alpha_2}{\alpha_3}
					   {\beta_1}{\beta_2}{\beta_3}
		\equiv
		\int_{-\infty}^{\eta_\ast} \frac{\d \eta}{\eta^4}
		\int_{-\infty}^{\eta} \frac{\d \tau}{\tau^4}
		(1 - \im \alpha_1 \eta + \alpha_2 \eta^2 + \im \alpha_3 \eta^3)
		\e{\im \alpha_1 \eta}
		\\ \mbox{} \qquad \qquad \qquad \times
		(1 - \im \beta_1 \tau + \beta_2 \tau^2 + \im \beta_3 \tau^3)
		\e{\im \beta_1 \tau} .
		\label{eq:three-ei-variant}
	\end{eqnarray}
	In order to evaluate the loop process, we shall need an expression for
	this integral in closed form. This can be obtained by iteratively
	integrating by parts, which produces a series containing
	``fast instabilities'' associated with negative powers of
	$\eta$, together with logarithmic contributions from integrals of
	the form $\int_{-\infty}^{\eta_\ast} \d \zeta \, \zeta^{-1}
	\e{\im \beta \zeta}$. This structure is the same as the one
	which we encountered when studying time integrals
	in \S\S\ref{sec:loop-spectrum}--\ref{sec:loop-bispectrum},
	although the result is slightly more complicated. This extra
	complexity stems from the fact that Eq.~\eref{eq:three-ei-variant}
	contains more negative powers of $\tau$ and $\eta$
	than (for example) Eq.~\eref{eq:time-two-asymptotic}
	and therefore requires more integrations by parts to separate
	its asymptotic dependence on $\eta_\ast$, and also because the
	integrands contain more terms anyway.
	Fewer terms are present when dealing with the gauge field
	because of its conformal
	invariance, which gives the associated gauge field wavefunctions
	a simpler temporal dependence than the minimally coupled scalar field.
	
	The time dependence which arises from Eq.~\eref{eq:three-ei-variant}
	is somewhat complicated to express. One obtains
	\begin{eqnarray}
		\fl\nonumber
		\threeEiVariant{\alpha_1}{\alpha_2}{\alpha_3}
					   {\beta_1}{\beta_2}{\beta_3} =
		\left(
			\frac{\sigma_1}{\eta_\ast^6} - \im \frac{\sigma_2}{\eta_\ast^5}
			+ \frac{\sigma_3}{\eta_\ast^4} - \im \frac{\sigma_4}{\eta_\ast^3}
			+ \frac{\sigma_5}{\eta_\ast^2} + \im \frac{\sigma_6}{\eta_\ast}
		\right) \e{\im \delta \eta_\ast}
		\\ \nonumber \mbox{}
		+ \delta \sigma_6 \int_{-\infty}^{\eta_\ast} \frac{\d \eta}{\eta}
		\e{\im \delta \eta}
		+ \left( - \im \frac{\rho_1}{\eta_\ast^3}
			- \frac{\rho_2}{\eta_\ast^2} - \im \frac{\rho_3}{\eta_\ast}
		\right) \e{\im \alpha_1 \eta_\ast}
		\int_{-\infty}^{\eta_\ast} \frac{\d \eta}{\eta} \e{\im \beta_1 \eta}
		\\ \mbox{}
		- ( \alpha_1 \rho_3 + \alpha_3 \lambda_4)
		\int_{-\infty}^{\eta_\ast} \frac{\d \eta}{\eta}
		\e{\im \alpha_1 \eta} \int_{-\infty}^{\eta} \frac{\d \tau}{\tau}
		\e{\im \beta_1 \tau} .
		\label{eq:three-ei-asymptotic}
	\end{eqnarray}
	The parameters appearing in this expression involve certain combinations
	of the $\alpha_i$ and $\beta_j$, which are described by the choices
	\begin{equation}
		\fl
		\lambda_1 \equiv \frac{1}{3}; \quad
		\lambda_2 \equiv \beta_1 \lambda_1; \quad
		\lambda_3 \equiv \beta_2 + \beta_1 \lambda_2; \quad
		\mbox{and} \quad
		\lambda_4 \equiv \beta_3 + \beta_1 \lambda_3;
	\end{equation}
	\begin{equation}
		\fl
		\gamma_1 \equiv \lambda_1 ; \quad
		\gamma_2 \equiv \lambda_2 + \alpha_1 \lambda_1 ; \quad
		\gamma_3 \equiv \lambda_3 + \alpha_2 \lambda_1 - \alpha_1 \lambda_2 ;
	\end{equation}
	\begin{equation}
		\fl
		\gamma_4 \equiv \alpha_1 \lambda_3 + \alpha_2 \lambda_2 - \alpha_3
			\lambda_1; \quad
		\gamma_5 \equiv \alpha_3 \lambda_2 + \alpha_2 \lambda_3; \quad
		\mbox{and} \quad
		\gamma_6 \equiv \alpha_3 \lambda_3 .
	\end{equation}
	We also define a combination $\delta$,
	\begin{equation}
		\delta \equiv \alpha_1 + \beta_1 .
	\end{equation}
	In terms of these combinations, the $\rho_i$ satisfy
	\begin{equation}
		\rho_1 \equiv \frac{\lambda_4}{3} ; \quad
		\rho_2 \equiv \frac{\alpha_1 ( \lambda_4 - \rho_1 )}{2} ; \quad
		\rho_3 \equiv \alpha_2 \lambda_4 + \alpha_1 \rho_2 .
	\end{equation}
	We now have all the ingredients necessary to define the
	$\sigma_i$:
	\begin{equation}
		\sigma_1 \equiv \frac{\gamma_1}{6} ; \quad
		\sigma_2 \equiv \frac{\gamma_2 - \delta \sigma_1}{5} ; \quad
		\sigma_3 \equiv \frac{\gamma_3 + \delta \sigma_2}{4} ;
	\end{equation}
	\begin{equation}
		\sigma_4 \equiv \frac{\gamma_4 - \delta \sigma_3 + \rho_1}{3} ;
			\quad
		\sigma_5 \equiv \frac{\gamma_5 + \delta \sigma_4 - \rho_2}{2} ;
			\quad
		\mbox{and}
	\end{equation}
	\begin{equation}
		\sigma_6 \equiv \gamma_6 + \delta \sigma_5 - \rho_3 .
	\end{equation}
	
	Let us focus first on the infra-red region, where the momentum
	circulating in the loop is small compared to the momenta
	$k_1 = k_2 = |\vect{k}|$ on the external legs. In this limit, we can
	take $q \approx 0$ and $r \approx k$.
	Although Eq.~\eref{eq:three-ei-asymptotic} contains a large number of
	primitive fast divergences (up to and including
	$\eta_\ast^{-6}$ in the pure power-law sector, and up to and including
	$\eta_\ast^{-2}$ when multiplied by additional logarithmic divergences,
	bearing in mind that purely imaginary divergences cancel out of the
	final answer), these all cancel in both the ultra-violet and infra-red.
	The leading time dependence instead comes from the double
	exponential integral. This can be written
	\begin{equation}
		\fl
		\int_{-\infty}^{\eta_\ast} \frac{\d \eta}{\eta} \e{\im \alpha_1 \eta}
		\int_{-\infty}^{\eta} \frac{\d \tau}{\tau} \e{\im \beta_1 \tau}
		= \frac{1}{2} \ln^2 |\alpha_1 \eta_\ast| +
		  \Big( \EulerMascheroni + \ln \left| \frac{\alpha_1}{\beta_1}
			\right| \Big) \ln |\alpha_1 \eta_\ast| + f( \alpha_1 / \beta_1) ,
	\end{equation}
	where $f(x)$ is a complicated function of its argument which can be
	given in closed form in terms of the Euler--Mascheroni constant when
	$x=1$. We will focus only on the leading divergence as $\eta_\ast
	\rightarrow 0$. Keeping only this term, one finds that the
	loop corrected power spectrum takes the form
	\begin{equation}
		P(k) = P_{0 \ast}(k) \left(
			1 +
			\frac{2}{9} \left( \frac{V_\ast'''}{2\pi} \right)^2
			\frac{N_\ast^2}{H_\ast^2} \ln k \ell 
		\right) ,
		\label{eq:v-loop-ir}
	\end{equation}
	where $P_{0}(k)$ is the tree-level power spectrum
	and $\ell$ is an infra-red regulator which has been introduced to
	prevent a divergence in the momentum integral. As described in
	Refs.~\cite{Sloth:2006az,Sloth:2006nu,Lyth:2007jh,Seery:2007we,
	Seery:2007wf,Bartolo:2007ti}, this regulator can be understood as the
	size of a comoving ``box'' in which we perform the quantum field
	theory calculation. This box must be chosen to be sufficiently large
	that the region of space for which we wish to obtain a prediction
	can fit comfortable inside it, but small enough that all fluctuations
	remain perturbatively small. Eq.~\eref{eq:v-loop-ir} is identical
	with Eq.~(1) of Ref.~\cite{Bartolo:2007ti}.
	
	In the opposite limit, where very hard virtual quanta are
	circulating in the loop, $q$ and $r$ are both much larger than
	$k$. In this limit, one finds that the loop correction has the form
	\begin{equation}
		P(k) = P_{0 \ast}(k) \left\{
			1 - \frac{16}{27} \frac{\ln (\Lambda'/H_\ast)}
			{H_\ast^2} \left( \frac{V_\ast'''}{2\pi} \right)^2
		\right\} .
	\end{equation}
	
	\section*{References}
	
	\providecommand{\href}[2]{#2}\begingroup\raggedright\endgroup

\end{fmffile}
\end{document}